\documentclass[a4paper,fleqn,usenatbib,useAMS]{mnras}

\usepackage[T1]{fontenc}
\usepackage{ae,aecompl}

\usepackage{graphicx}	% Including figure files
\usepackage{stackengine}
\usepackage{epstopdf}
\usepackage{savesym}
\usepackage{amsmath}
\savesymbol{iint}
\savesymbol{iiint}
\usepackage{txfonts}
\restoresymbol{TXF}{iint}
\restoresymbol{TXF}{iiint}
\usepackage{fix-cm}
\usepackage{subfig}
\usepackage[]{units}
\usepackage{textcomp}

\title[Zonal flow evolution and overstability in discs]{Zonal flow evolution and overstability in accretion discs}

\author[R. Vanon \& G.I. Ogilvie]{
R. Vanon\thanks{E-mail: \href{mailto:rv288@cam.ac.uk}{rv288@cam.ac.uk}}
and G. I. Ogilvie
\\
Department of Applied Mathematics and Theoretical Physics, University of Cambridge, Centre for Mathematical Sciences, Wilberforce Road, Cambridge\\
CB3 0WA, UK
}

\date{Accepted XXX. Received YYY; in original form ZZZ}

\pubyear{2016}

\begin{document}
\label{firstpage}
\pagerange{\pageref{firstpage}--\pageref{lastpage}}
\maketitle

\begin{abstract}
  This work presents a linear analytical calculation on the stability and evolution of a compressible, viscous self-gravitating (SG) Keplerian disc with both horizontal thermal diffusion and a constant cooling timescale when an axisymmetric structure is present and freely evolving. The calculation makes use of the shearing sheet model and is carried out for a range of cooling times. 
  Although the solutions to the inviscid problem with no cooling or diffusion are well known, it is non-trivial to predict the effect caused by the introduction of cooling and of small diffusivities; this work focuses on perturbations of intermediate wavelengths, therefore representing an extension to the classical stability analysis on thermal and viscous instabilities. For density wave modes the analysis can be simplified by means of a regular perturbation analysis; considering both shear and thermal diffusivities, the system is found to be overstable for intermediate and long wavelengths for values of the Toomre parameter $Q\lesssim2$; a non-SG instability is also detected for wavelengths $\gtrsim 18H$, where $H$ is the disc scale height, as long as $\gamma \lesssim 1.305$. 
  The regular perturbation analysis does not however hold for the entropy and potential vorticity slow modes as their ideal growth rates are degenerate. To understand their evolution, equations for the axisymmetric structure's amplitudes in these two quantities are analytically derived and their instability regions obtained. The instability appears boosted by increasing the value of the adiabatic index and of the Prandtl number, while it is quenched by efficient cooling. 
\end{abstract}


%%%%%%%%%%%%%%%%%%%%%%%%%%%%%%%%%%%%%%%%%%%%%%%%%%

%%%%%%%%%%%%%%%%% BODY OF PAPER %%%%%%%%%%%%%%%%%%

\section{Introduction}
Accretion discs are subject to an assortment of instabilities; two of the most widely studied instances are the classical thermal and viscous instabilities \citep[e.g.][]{Pringleetal1973, LightmanEardley1974, ShakuraSunyaev1976, Pringle1977, LivioShaviv1977, Piran1978, Pringle1981}. Their existence depends on assumptions about how the angular momentum transport and dissipation are modelled, which distinguishes them from more fundamental dynamical instabilities such as the Magnetorotational Instability, the Gravitational Instability and the Vertical Shear Instability.

In a Keplerian disc of surface density $\Sigma$ and angular frequency $\Omega$ which is in thermal equilibrium, the heating and cooling rates $\mathcal{H}$ and $\mathcal{C}$ are equal and are given by

\begin{equation}
  \mathcal{H} = \frac{9}{4} \nu \Sigma \Omega^2 \propto \alpha T_c \Sigma \Omega
\end{equation}
\begin{equation}
  \mathcal{C} = 2 \sigma T_{\mathrm{eff}}^4 \propto \frac{T_c^4}{\tau},
\end{equation}
where $\nu = \alpha c_{\mathrm{iso}} H$ is the kinematic viscosity (with $c_\mathrm{iso}\propto T_c^{1/2}$ and $H=c_\mathrm{iso}/\Omega$ being the isothermal sound speed and the disc scale height), $\sigma$ is the Stefan-Boltzmann constant, $\tau$ (here assumed $\gg 1$) is the optical thickness and $T_c$ and $T_{\mathrm{eff}}$ are the central and effective temperatures of the disc.

As both $\alpha$ and $\tau$ are potentially functions of $T_c$, the disc is thermally unstable to perturbations in $T_c$ if
\begin{equation}
  \left. \frac{\partial \ln \mathcal{H}}{\partial \ln T_c}\right\rvert_\Sigma > \left. \frac{\partial \ln \mathcal{C}}{\partial \ln T_c}\right\rvert_\Sigma,
\end{equation}
as it would lead to runaway heating (cooling) for an upward (downward) temperature perturbation. In the above criterion the surface density $\Sigma$ is held constant as changes in temperature happen on a much shorter timescale than changes in $\Sigma$ due to the thermal timescale $\tau_{\mathrm{th}}$ being given by

\begin{equation}
  \tau_{\mathrm{th}} \simeq  \left(\frac{H}{R}\right)^2 \tau_{\mathrm{visc}},
\end{equation}
with $\tau_{\mathrm{visc}}$ representing the viscous timescale and $H/R \ll 1$ for a thin disc.

The $\alpha$ model of accretion discs \citep{ShakuraSunyaev1973} predicts the disc to be thermally unstable in the inner regions (where the radiation pressure dominates), although it is uncertain whether the thermal instability predicted by the $\alpha$ model takes place in real discs, with some observations seeming to have proven otherwise \citep[eg.][]{GierlinskiDone2004, Doneetal2007}; a competing model exists (dubbed the $\beta$ model) where the stress is proportional to the gas pressure, rather than the total pressure as in the $\alpha$ model. This produces a thermally stable disc \citep{SakimotoCoroniti1981, StellaRosner1984, Merloni2003}. Moreover, the $\alpha$ model neglects other effects such as heating from MRI-induced turbulence \citep[eg.][]{Hiroseetal2009} and heat transport within the disc.

A disc is said to be viscously unstable if a perturbation $\delta \mu$ applied to the dynamic viscosity $\mu = \nu \Sigma$ grows. Substituting this perturbation into the equation of diffusive disc evolution 

\begin{equation}
  \frac{\partial \Sigma}{\partial t} = \frac{3}{r} \frac{\partial}{\partial r} \left[r^{1/2} \frac{\partial}{\partial r} \left(\nu \Sigma r^{1/2}\right) \right]
\end{equation}
gives 

\begin{equation}
  \frac{\partial}{\partial t} (\delta \mu) = \frac{\partial \mu}{\partial \Sigma} \frac{3}{r} \frac{\partial}{\partial r} \left[r^{1/2} \frac{\partial}{\partial r}\left(r^{1/2} \delta \mu\right)\right],
\end{equation}
with an instability being triggered if the diffusion coefficient is negative. This implies the viscous instability criterion to be \citep{LightmanEardley1974}

\begin{equation} \label{eq:visc-inst-criterion}
  \frac{\partial (\nu \Sigma)}{\partial \Sigma} < 0,
\end{equation}
with the derivative being taken at constant $r$, and under the assumption of both thermal balance and hydrostatic equilibrium.

The classical approach does however have limitations, the most notable of which being the consideration of long wavelength perturbations obeying $H \ll \lambda_{\mathrm{pert}} \ll R_0$ only, which in turn allows the thermal and viscous instabilities to be distinct. A more general analysis can be conducted by considering perturbations of wavelength $\lambda_{\mathrm{pert}} \sim H$; in this case the previously existing structure in the density also develops a significant perturbation in the azimuthal component of the velocity, therefore becoming a zonal flow, which modifies the shear rate from its Keplerian value. This more generic analysis can be used to study the stability of the slow modes and establish whether zonal flows grow or decay as a result of non-ideal effects such as viscous interactions, cooling and heating, as well as the coupling between the modes. 

Zonal flows -- axisymmetric shear flows consisting of parallel bands -- represent an equilibrium solution to the equations governing the evolution of an accretion disc's flow, involving a geostrophic balance between the Coriolis force and the pressure gradient. This can however be unstable under certain conditions, in which case the flow can undergo a Kelvin-Helmholtz (or Rossby wave) instability \citep{VanonOgilvie2016}. Zonal flows have been observed to persist in certain conditions; one such example is 3D simulation of MHD-turbulent discs modelled using the shearing box approximation \citep{Johansenetal2009, Simonetal2012, KunzLesur2013, BaiStone2014}. In this scenario zonal flows are seen to exhibit larger amplitudes and longer lifetimes for larger boxes \citep{BaiStone2014}, although the correlation between lifetime and box size doesn't appear to hold for boxes of very small size \citep{Johansenetal2009}. 2D shearing sheet hydrodynamical simulations of accretion discs have also encountered persistent zonal flows -- albeit with a finite lifetime -- that are found to be unstable to the formation of long-lived vortices \citep{UmurhanRegev2004, JohnsonGammie2005, Lithwick2007, Lithwick2009}. This is regardless of the modest Reynolds numbers achievable in simulations compared to those describing real discs. The emergence and survival of zonal flows in both hydrodynamical and MHD simulations could be crucial in the context of planetesimal growth within protoplanetary discs. Their presence can in fact alter the coupling between the disc gas and the planetesimals \citep{Weidenschilling1977}, helping the latter to overcome their inward migration due to gas drag \citep{KlahrLin2001, FromangNelson2005, Katoetal2009} when planetesimals reach the `metre-sized barrier', while at the same time promoting their growth. 

A disc can also be viscously unstable to axisymmetric oscillations, as first described by \citet{Kato1978}. He found that if a disc's turbulent viscosity coefficient increases in compressive motions this would generate a larger amount of thermal energy, therefore leading to the growth of the axisymmetric oscillations, in a mechanism that is comparable to the generation of nuclear energy driving stellar pulsations. Furthermore, \citet{Kato1978} found -- by means of a local stability analysis -- that said oscillations can undergo an overstability if the viscosity coefficient increases sufficiently rapidly with the surface density.
Since the seminal work by \citet{Kato1978}, the viscously overstable regime has been applied to the $\alpha$-disc model \citep{Blumenthaletal1984} -- where the oscillations were found to become viscously overstable if the value of $\alpha$ exceeds a critical value -- and analysed in both linear and non-linear regimes in planetary rings and gaseous disc contexts \citep[eg.][]{KatoFukue1980, Borderiesetal1985, PapaloizouStanley1986, Katoetal1988, PapaloizouLin1988, SchmitTscharnuter1999, LatterOgilvie2006}. A fresh look is taken at the topic of overstability in this analysis, also considering how this is affected by self-gravity.

This work presents an analytical calculation of the evolution and stability of the solutions to a compressible, viscous self-gravitating Keplerian disc with horizontal thermal diffusion when an axisymmetric structure is present. The disc, which is modelled using the 2D shearing sheet approximation, also possesses a constant $\beta$ cooling, with a range of values used in the analysis. The work focuses on perturbations of wavelengths $\lambda_{\mathrm{pert}} \sim H$, rather than $H \ll \lambda_{\mathrm{pert}} \ll R_0$ as in the classical works dealing with thermal and viscous instabilities; our work therefore represents an extension of the classical theory of said instabilities. The paper is arranged as follows: Section~\ref{sec:model} serves as an introduction to the shearing sheet model which is employed in this analysis, as well as the full non-linear, viscous equations governing the system described. Section~\ref{sec:evolution} introduces the axisymmetric structure and the equations describing its temporal evolution; it also analyses the evolution and stability of both density waves and slow modes. The work terminates in Section~\ref{sec:conclusions}, where the conclusion drawn from the results are presented.

\section{Model} \label{sec:model}
The work presented in this paper is based on the local unstratified shearing sheet model, whose first use was by \citet{GoldreichLynden-Bell1965} in the context of galactic discs. This consists in drawing a sheet of small dimensions compared to the disc size centred at a fiducial radius $R_0$ (i.e. $L_x, L_y \ll R_0$, where $L_x$ and $L_y$ are the radial and azimuthal dimension of the chosen sheet). The frame of reference of the sheet, which is of a Cartesian nature, co-rotates with the disc at an angular frequency $\boldsymbol\Omega = \Omega \mathbf{e}_z$, with $\mathbf{e}_z$ being the unit vector normal to the sheet; in the chosen frame of reference, the continuity and Navier-Stokes equations for a viscous, compressible fluid are given by

\begin{equation}
  \partial_t \Sigma + \nabla \cdot (\Sigma \boldsymbol{\varv}) = 0,
\end{equation}
\begin{equation}
  \partial_t \boldsymbol{\varv} + \boldsymbol{\varv} \cdot \nabla \boldsymbol{\varv} + 2 \boldsymbol{\Omega} \times \boldsymbol{\varv} = - \nabla \Phi - \nabla \Phi_{\mathrm{d,m}} - \frac{1}{\Sigma} \nabla P + \frac{1}{\Sigma} \nabla \cdot \mathbf{T},
\end{equation}
where $\Sigma$ is the surface density of the disc, $\boldsymbol{\varv}$ is the velocity of the flow, $\Phi=-q\Omega^2 x^2$ is the effective tidal potential (with $q=-\mathrm{d}\ln \Omega/\mathrm{d} \ln r$ representing the dimensionless shear rate, its value being $q=3/2$ for a Keplerian disc), $\Phi_{\mathrm{d,m}}$ is the disc potential evaluated at its mid-plane, $P$ is the 2-dimensional pressure and $\mathbf{T} = 2 \mu_s \mathbf{S} + \mu_b \left(\nabla \cdot \boldsymbol\varv\right) \mathbf{I}$ is the viscous stress tensor, with $\mathbf{S} = \frac{1}{2}\left[\nabla \boldsymbol{\varv} + (\nabla \boldsymbol{\varv})^T\right] - \frac{1}{3}(\nabla \cdot \boldsymbol{\varv}) \mathbf{I}$ being the traceless shear tensor, $\mu=\Sigma \nu$ the dynamic viscosity ($\mu_s$ and $\mu_b$ being the shear and bulk dynamic viscosities, respectively), $\nu$ the kinematic viscosity and $\mathbf{I}$ the unit tensor.

The quantity $h = \ln \Sigma + \mathrm{const.}$ is introduced, which turns the continuity equation into

\begin{equation}
  \partial_t h + \boldsymbol{\varv} \cdot \nabla h + \nabla \cdot \boldsymbol{\varv} = 0,
\end{equation}
while the disc potential can be readily evaluated at the disc's mid-plane in Fourier space by means of Poisson's equation $\nabla^2 \Phi_\mathrm{d} = 4 \pi G \Sigma \delta(z)$, its form being described by

\begin{equation}
  \widetilde{\Phi}_{\mathrm{d,m}} = - \frac{2\pi G\widetilde{\Sigma}}{\sqrt{k_x^2 + k_y^2}},
\end{equation}
where $G$ is the gravitational constant and $k_x$ and $k_y$ are the radial and azimuthal components of the wave vector $\mathbf{k}$.

Another crucial equation in the setup described is that for the temporal evolution of the specific internal energy $e$, which is given by

\begin{multline}
  \partial_t e + \boldsymbol{\varv} \cdot \nabla e = - \frac{P}{\Sigma} \nabla \cdot \boldsymbol{\varv} + 2 \nu_s \mathrm{\mathbf{S}}^2 + \nu_b (\nabla \cdot \boldsymbol{\varv})^2 + \frac{1}{\Sigma}\nabla \cdot (\nu_t \Sigma \nabla e) \\- \frac{1}{\tau_c} (e-e_{\mathrm{irr}}),
\end{multline}
where $\nu_b$ and $\nu_s$ are the bulk and shear kinematic viscosities, $\nu_t$ the (horizontal) thermal diffusion, $\tau_c$ the (constant) cooling timescale and $e_{\mathrm{irr}}$ the equilibrium specific internal energy to which the disc would relax if it were not viscously heated. The Prandtl number is defined as 

\begin{equation}
  \mathrm{Pr} = \frac{\nu_s}{\nu_t}.
\end{equation}

The analysis conducted in this paper will also make use of two quantities which are material invariants in ideal conditions (i.e. in the absence of diffusivities and cooling): potential vorticity $\zeta$ (PV) and the dimensionless specific entropy $s$, whose forms are given by

\begin{equation}
  \zeta = \frac{2\Omega + \left(\nabla \times \boldsymbol{\varv}\right)_z}{\Sigma},
\end{equation}
\begin{equation}
  s = \frac{1}{\gamma} \ln P - \ln \Sigma,
\end{equation}
where $\gamma$ represents the adiabatic index. The pressure $P$ is given in terms of the specific internal energy $e$ by 

\begin{equation}
  P = (\gamma-1) \Sigma e.
\end{equation}
This allows us to evaluate the pressure gradient term in the momentum equation as

\begin{equation}
  \frac{\nabla P}{\Sigma} = (\gamma-1) \left( \nabla e + e \nabla h\right).
\end{equation}

The background state of the system is described by $\Sigma=\Sigma_0$, $\boldsymbol\varv_0=(0,-q \Omega x,0)^T$ and by an internal energy per unit mass $e=e_0=c_s^2/(\gamma(\gamma-1))$, where $c_s$ is the adiabatic sound speed; the introduction of an internal energy induced by external irradiation $e_{\mathrm{irr}}$ acts as a buffer in the thermal balance of the system. Whereas in its absence thermal balance can only be achieved with one combination of cooling time and shear viscosity, the assumption that $e_{\mathrm{irr}} \geq 0$ allows us to explore multiple permutations of the two parameters to gauge their effect on disc stability. The thermal balance of the background state is given by

\begin{equation} \label{eq:therm-eq-1}
  e_0 = e_{\mathrm{irr}} + e_0 \alpha_s (\gamma-1) q^2 \Omega \tau_c.
\end{equation}
It is possible to identify the quantity 

\begin{equation}
  f_\mathrm{visc} = \alpha_s (\gamma-1) q^2 \Omega \tau_c, 
\end{equation} 
which represents the fraction of viscously generated heat, with $e_\mathrm{irr}=0$ (ie. disc being entirely viscously heated) yielding the maximum value of $f_\mathrm{visc}=1$.
Equation~\ref{eq:therm-eq-1}, under the assumption $e_{\mathrm{irr}}\geq 0$, implies that

\begin{equation} \label{eq:therm-eq-condition}
  \alpha_s \tau_c \leq \frac{1}{q^2 \Omega (\gamma-1)},
\end{equation}
where $\alpha_s = \nu_s \left( \frac{\gamma \Omega}{c_s^2}\right)$ is a dimensionless viscosity parameter, which defines our ranges of shear viscosity and cooling timescale ranges for a specific dimensionless shear rate and adiabatic index. 

The background state is then perturbed such that $\boldsymbol{\varv} = \boldsymbol{\varv}_0 + \boldsymbol{\varv}^\prime$ (with $\boldsymbol{\varv}^\prime = (u^\prime,\varv^\prime,0)^T$), etc. This yields the following set of linearised equations describing the temporal evolution of the disturbance:

\begin{equation} \label{eq:nonlin_h}
  \partial_t h^\prime = - \partial_x u^\prime ,
\end{equation}
\begin{multline} \label{eq:nonlin_u}
  \partial_t u^\prime - 2\Omega \varv^\prime = - \partial_x \Phi^\prime_{\mathrm{d,m}} - (\gamma-1) \left[ \partial_x e^\prime + e_0 \partial_x h^\prime \right] \\ + \left(\nu_b + \frac{4}{3}\nu_s\right) \partial_x^2 u^\prime ,
\end{multline}
\begin{equation} \label{eq:nonlin_v}
  \partial_t \varv^\prime + (2-q) \Omega u^\prime = \nu_s \partial_x^2 \varv^\prime - \nu_s q\Omega \partial_x h^\prime,
\end{equation}
\begin{equation} \label{eq:nonlin_e}
  \partial_t e^\prime = -(\gamma-1) e_0 \partial_x u^\prime - 2 \nu_s q\Omega \partial_x \varv^\prime + \nu_t \partial_x^2 e^\prime - \frac{1}{\tau_c}(e^\prime - e_{\mathrm{irr}}),
\end{equation}
with the analysis being based on the assumptions of $\tau_c=\mathrm{const}$ and $\nu_i = \mathrm{const}$. It is worth noting that the assumption of constant diffusivities made can potentially affect the stability properties of the model described.

As further explored in Section~\ref{sec:evolution}, the solutions to the above equations -- which are either density waves (DWs) or non-oscillating structures in the entropy and potential vorticity -- are deeply influenced by the viscosity and thermal diffusivity values, as well as the effectiveness of the imposed cooling. Depending on their combined effects, the solutions to the problem can be either damped, exponentially growing or overstable (i.e. growing oscillations).

\section{Evolution} \label{sec:evolution}
The system admits axisymmetric, sinusoidal standing-wave solutions of the form 

\begin{align} \label{eq:ad-structure}
  h^\prime(x,t) & = A_h(t) \cos(kx) \nonumber \\
  u^\prime(x,t) & = A_u(t) \sin(kx) \nonumber \\
  \varv^\prime(x,t) & = A_\varv(t) \sin(kx) \nonumber \\
  e^\prime(x,t) & = e_0 A_e(t) \cos(kx),
\end{align}
where $A_h$, $A_u$, $A_\varv$ and $A_e$ represent the amplitudes in the respective quantities and $k>0$ is the wavenumber of the above structure.

It is possible to obtain a set of equations describing the temporal evolution of the axisymmetric structure by applying its form outlined above into the linearised equations describing the system (Equations~\ref{eq:nonlin_h}--\ref{eq:nonlin_e}):
\begin{equation} \label{eq:str-h}
  \partial_t A_h = - k A_u,
\end{equation}
\begin{equation} \label{eq:str-u}
  \partial_t A_u - 2\Omega A_\varv = - 2\pi G \Sigma_0 A_h + \frac{c_s^2 k (A_e + A_h)}{\gamma} - \left(\gamma_b + \frac{4}{3}\gamma_s\right) A_u,
\end{equation}
\begin{equation} \label{eq:str-v}
  \partial_t A_\varv + (2-q)\Omega A_u = - \gamma_s A_\varv + \gamma_s \frac{q\Omega}{k} A_h,
\end{equation}
\begin{equation} \label{eq:str-e}
  \partial_t A_e = - (\gamma-1)k A_u - \gamma_s \frac{2q\Omega}{k e_0} A_\varv - \gamma_t A_e,
\end{equation}
where $\gamma_b = \nu_b k^2$, $\gamma_s = \nu_s k^2$ and $\gamma_t = \nu_t k^2 + 1/\tau_c$ are three damping coefficients. 

\begin{figure}
  \includegraphics[width=\columnwidth]{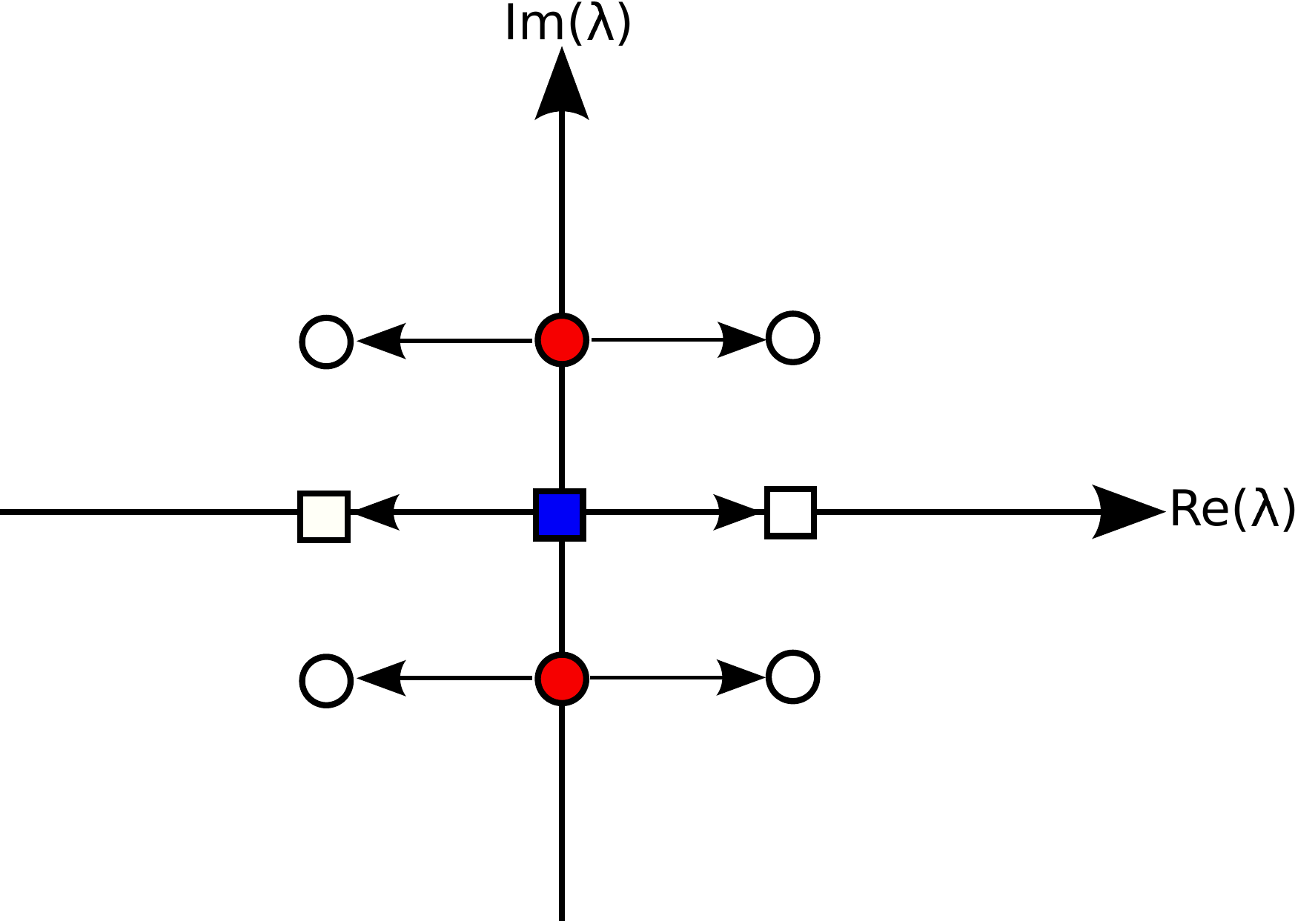}
  \caption{Graphic illustration of the possible solutions to the ideal (inviscid with no cooling; filled shapes) and full cases (empty shapes) in the Real--Imaginary growth rate plane. In the inviscid case all modes have $\mathrm{Re}(\lambda)=0$, with the potential vorticity (PV) and entropy modes, both having zero-frequency, being indistinguishable (blue square). As viscous terms and cooling are introduced, the modes acquire a non-zero real part to their growth rates; if $\mathrm{Re}(\lambda)<0$ viscosity acts to dampen disturbances, while if $\mathrm{Re}(\lambda)>0$ the entropy/PV modes (white squares) exhibit exponential growth while the density wave modes (white circles) are subject to overstability.}
  \label{fig:modes}
\end{figure}

If we assume that these equations have solutions of the form $\propto \mathrm{e}^{\lambda t}$, a quartic equation for the complex growth rate $\lambda$ can be determined, and its solutions analysed.
In the inviscid case with no cooling or diffusion, these will be
\begin{equation}
  \lambda_0 = 0, \, 0, \, \pm \, \mathrm{i} \, \omega_0 ,
\end{equation} 
where the zero subscript indicates ideal case considered, $\omega_0^2 = \kappa^2 - 2\pi G\Sigma_0 k + c_s^2 k^2$ is the square of the density wave frequency and $\kappa^2 = 2(2-q)\Omega^2$ is the epicyclic frequency squared. The two non-zero roots correspond to the density wave modes, while the zero roots correspond to the potential vorticity and entropy slow modes, as indicated in Figure~\ref{fig:modes} by the filled shapes. The density waves are stable for all $k$ values if $Q>1$, where $Q$ is the Toomre parameter -- which represents the strength of self-gravity (SG) within a disc, with $Q \lesssim 1$ causing the disc to be gravitationally unstable -- given by 

\begin{equation}
  Q \equiv \frac{c_s \kappa}{\pi G \Sigma_0}.
\end{equation}

Introducing the damping coefficients (assumed to be small, i.e. $\gamma_i \ll \Omega$) back into the picture gives a non-zero real part to all the modes' growth rates, as shown in Figure~\ref{fig:modes} by the empty shapes. If the newly acquired real part is negative, the damping coefficients have a stabilising effect on the modes, while if $\mathrm{Re}(\lambda)>0$ the modes exhibit exponential growth (entropy and PV modes) or viscous overstability (DW modes). 
Understanding how the introduction of the three diffusivities affects the values of the solutions is however non-trivial. It is expected that a regular perturbation analysis can be made for non-degenerate eigenvalues (i.e. for the density wave modes with $\lambda_0 = \pm \mathrm{i} \omega_0$), assuming the diffusivity values are small enough; in this case, the solutions to the full equations are
  
\begin{equation} \label{eq:lambda-linear}
  \lambda = \lambda_0 + \sum_{i=1}^{3} \gamma_i  \left( \frac{\partial \lambda}{\partial \gamma_i}\right) + \mathcal{O}\left(\gamma_i^2\right),
\end{equation}
where $\gamma_i$ can represent a bulk, shear or thermal damping coefficient, the latter also including effects due to cooling. 
In the degenerate case (i.e. entropy/PV modes with $\lambda_0 = 0$) it is however possible that a singular perturbation is necessary, meaning the solutions would not agree with the expression given by Equation~\ref{eq:lambda-linear}.

\subsection{Density wave modes}
The linearisation assumption is found to hold for density wave modes (i.e., the non-zero roots in the inviscid case), and the independent contributions to these modes from the damping coefficients are calculated using the eigenvalue problem; these are

\begin{gather}
  \left( \frac{\partial \lambda}{\partial \gamma_b}\right) = - \frac{1}{2}, \nonumber \\
  \left( \frac{\partial \lambda}{\partial \gamma_t}\right) = - \frac{k^2c_s^2 \left(\gamma-1\right)}{2\gamma\omega_0^2},\\
  \left( \frac{\partial \lambda}{\partial \gamma_s}\right) = \left[\left(\gamma-1\right)q^2 + 2\left(2-\gamma\right)q -2\right] \frac{\Omega^2}{\omega_0^2} - \frac{2}{3}. \nonumber
\end{gather}

\begin{figure}
  \includegraphics[width=\columnwidth]{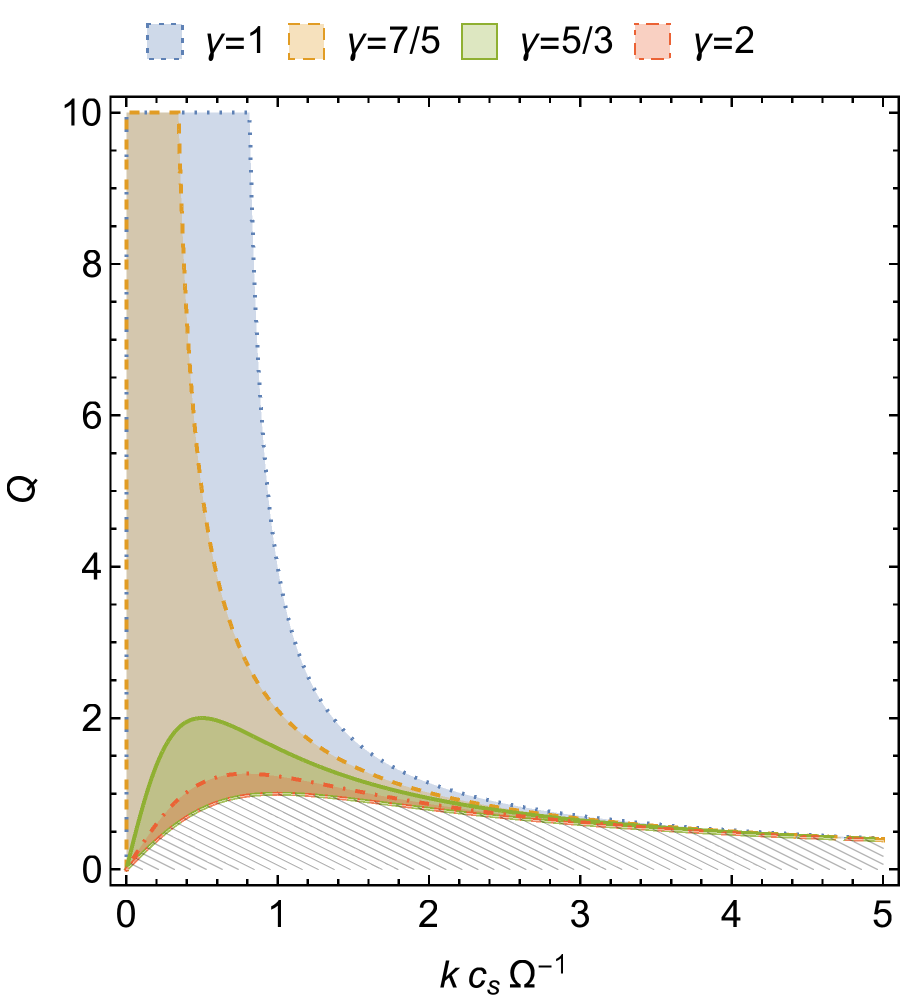}
  \caption{Stability of the density wave modes in the $k c_s/\Omega$ -- $Q$ plane under the influence of a shear viscosity alone, for various values of the adiabatic index $\gamma$, assuming the shear rate to be $q=3/2$. The shaded regions represent the parameter combinations for which a viscous overstability would ensue. While a viscous overstability can be triggered in a non-SG regime for $\gamma=1$ (blue, dotted region) and $\gamma=7/5$ (orange, dashed), increasing the value of $\gamma$ further to $\gamma=5/3$ (green, full) or $\gamma=2$ (red, dot-dashed) eliminates the high-$Q$ overstability region. Overstability in the latter two cases exists only for $Q\lesssim2$ for a broad range of wavenumbers, although the more unstable value appears to be $k c_s/\Omega \approx 0.5$. The hatched area represents the region of the parameter space where $\omega_0^2<0$ and the flow is therefore dynamically unstable. }
  \label{fig:overstability_alphas}
\end{figure}

\begin{figure*}
  \centering
  \begin{minipage}{.99\textwidth}    
    \subfloat[]{\includegraphics[width=.47\columnwidth]{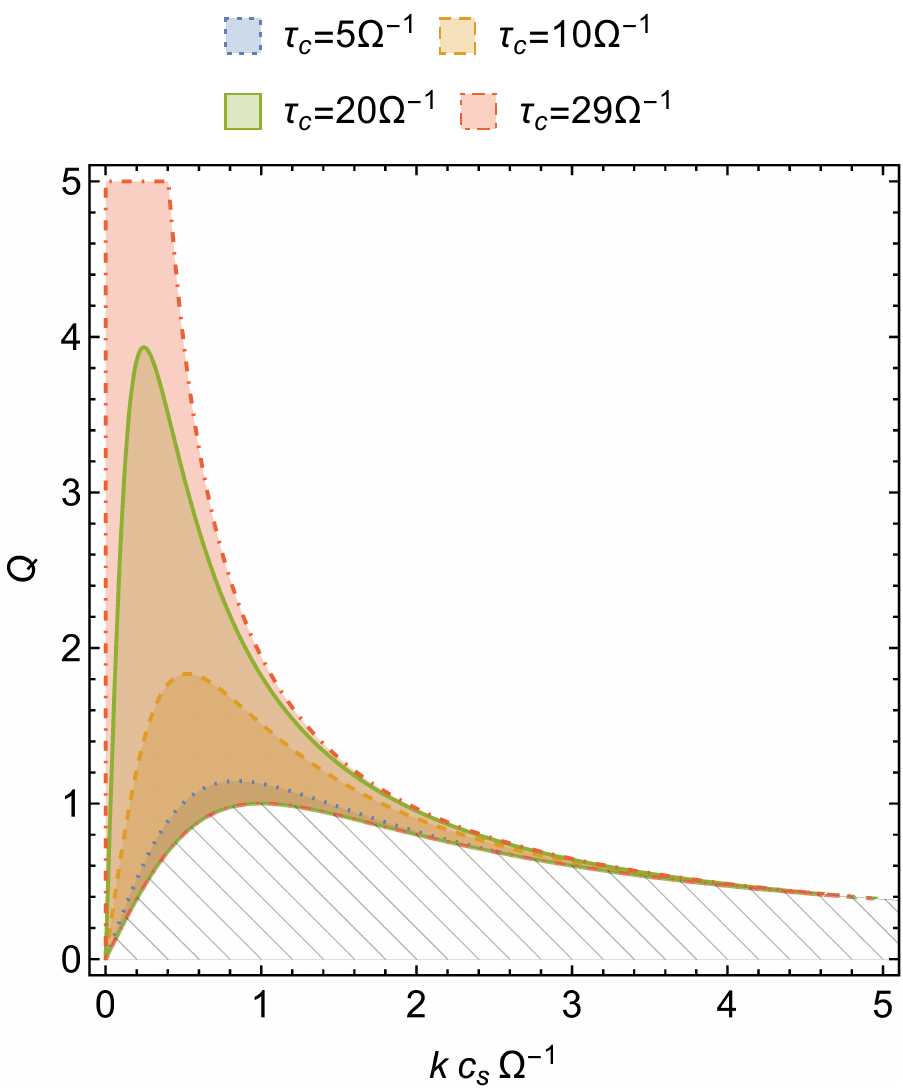}}
    \hspace*{.05\columnwidth}
    \subfloat[]{\includegraphics[width=.47\columnwidth]{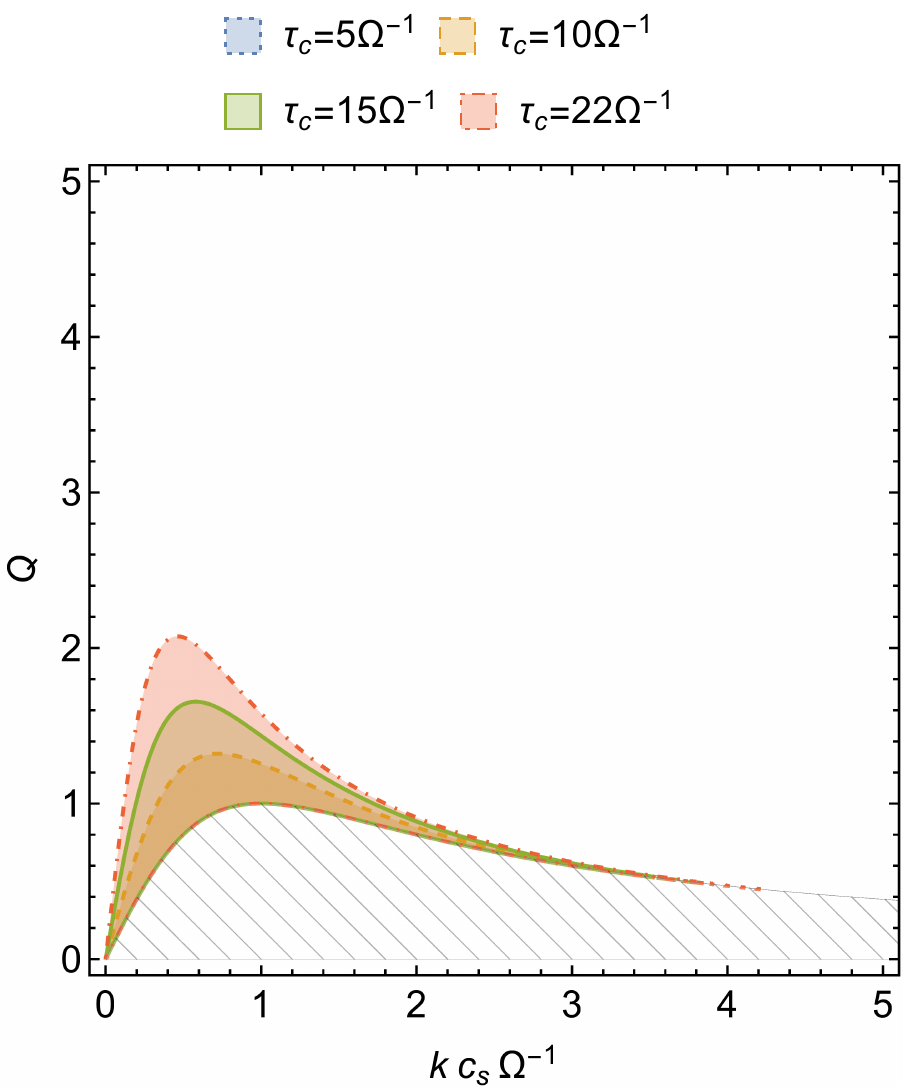}}
    \caption{Overstability regions under the influence of shear viscosity, thermal diffusion and cooling for $q=3/2$ and \textbf{(a)} $\gamma=1.3$ and \textbf{(b)} $\gamma=1.4$ (or $\gamma=7/5$). The analysis is carried out for a Prandtl number $\mathrm{Pr}=5$ (with $\alpha_s=0.05$) and various cooling timescales permitted by thermal balance. The $\gamma=1.3$ case retains a weak-SG/non-SG overstability at long wavelengths for the two longest cooling times: $\tau_c=20\Omega^{-1}$ (green, full lines) and $\tau_c=29\Omega^{-1}$ (red, dot-dashed). This is however not the case for the plot with $\gamma=1.4$ which only shows overstability for $Q\lesssim2$, as the value of $\gamma$ used in this case is larger than the predicted threshold value of $\gamma \simeq 1.305$. In both plots it's possible to see that cooling has a stabilising effect on the system, with shorter timescales progressively shrinking the overstability region. The hatched portion of the plot represents the region where the density wave frequency $\omega_0^2<0$ and the system is therefore gravitationally unstable to axisymmetric disturbances.}
    \label{fig:overstability_at_as}
  \end{minipage}
\end{figure*}

\begin{figure}
  \includegraphics[width=\columnwidth]{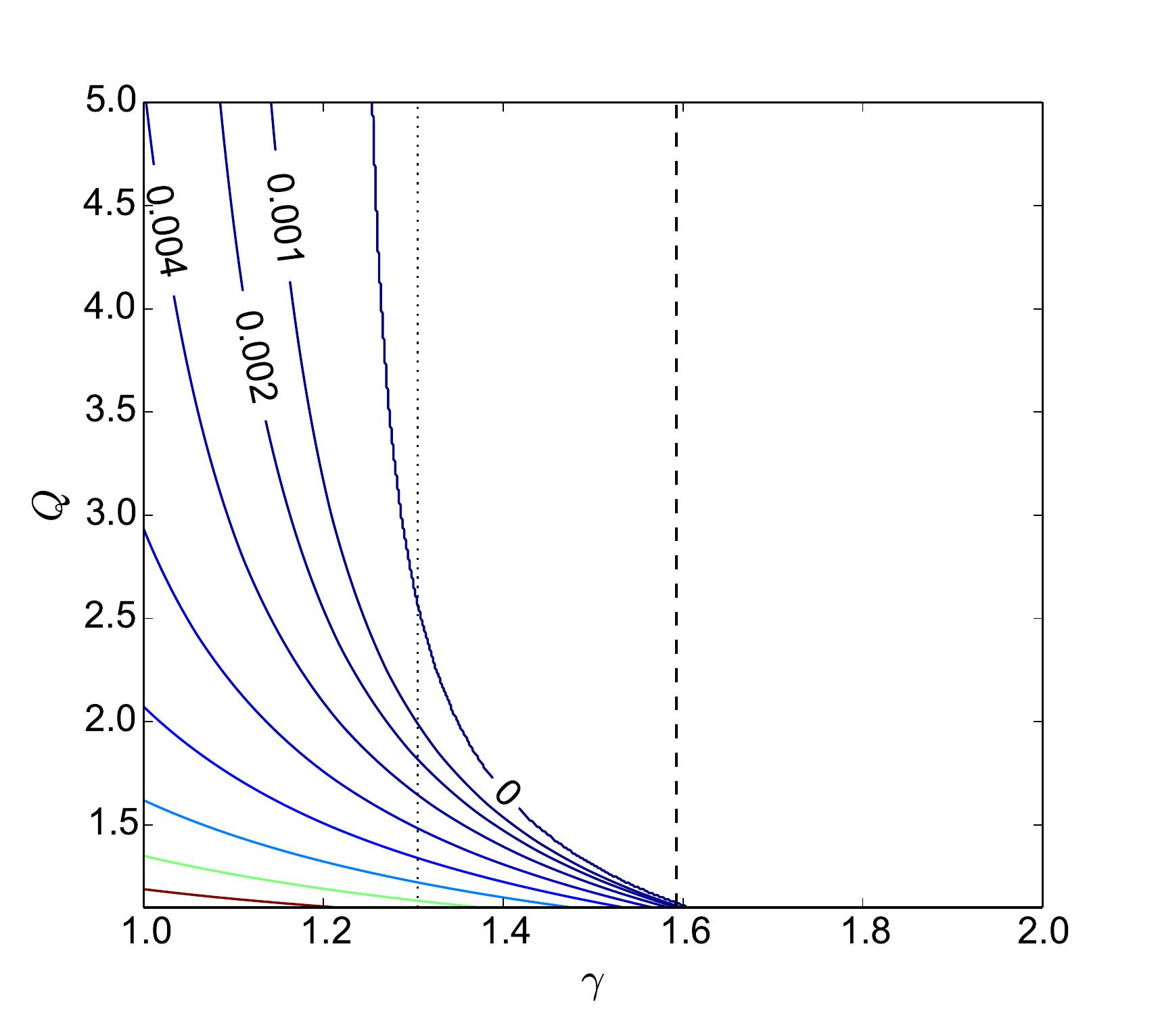}
  \caption{Overstability growth rates maximised over $k$ as a function of the adiabatic index $\gamma$ and Toomre parameter $Q$ for $q=3/2$ and a cooling time of $\tau_c=15\Omega^{-1}$. The values of the Prandtl number $\mathrm{Pr}=5$ and of the shear viscosity $\alpha_s=0.05$ match those employed in Figure~\ref{fig:overstability_at_as}. The vertical dashed line shows the largest value of $\gamma$ allowed by thermal balance ($\gamma \approx 1.6$), with larger values not permitted. The dotted vertical line represents the predicted threshold value of $\gamma\simeq 1.305$ above which a non-SG overstability cannot be achieved. As expected a large range of adiabatic index values offer unstable conditions when $Q\sim 1$, but only values of $\gamma \lesssim 1.25$ are overstable when $Q\sim 5$.}
  \label{fig:os_Q-vs-gamma}
\end{figure}

While it's clear to see that the contribution from the bulk viscosity is always negative, meaning it will always have a stabilising effect on the density wave modes, the situation is more intricate in the case of the shear viscosity and thermal diffusion. Should the contribution from a specific diffusivity type happen to be positive, it would imply that diffusivity type would act towards causing the density wave modes to be overstable. However an overstability is only reached if the total contribution $\sum_i \gamma_i  \left( \frac{\partial \lambda}{\partial \gamma_i}\right)$ is positive.

While the thermal diffusion also has a stabilising contribution when $\omega_0^2>0$ (where the flow is dynamically stable), for the shear viscosity the contribution is a more complicated expression which depends on $\gamma$ and $q$, as well as $k$ and $Q$. However it should be noted that the expression enclosed within square brackets in $\left( \frac{\partial \lambda}{\partial \gamma_s}\right)$ is positive for most realistic value combinations of $q$ and $\gamma$. The regions where overstability occurs when only shear viscosity is taken into account are shown in Figure~\ref{fig:overstability_alphas} for a range of values of the adiabatic index $\gamma$, assuming $q=3/2$.

$\gamma=1$ (blue, dotted region) and $\gamma=7/5$ (orange, dashed) produce an overstable region in the $k c_s/\Omega$--$Q$ plane that extends to arbitrarily high $Q$ for sufficiently large wavelengths ($\gtrsim 9 H$ and $\gtrsim 16 H$, respectively, where $H=c_\mathrm{iso}/\Omega$ is the scale height of the disc), as well as a low-$Q$ region traversing the whole range of $k c_s/\Omega$ considered, which is consistent with the result of \citet{LatterOgilvie2006}. When the value of $\gamma$ is further increased, the high-$Q$ region becomes stable, leaving only the low-$Q$ overstability region for $\gamma=5/3$ (green, full) and $\gamma=2$ (red, dot-dashed), which also appears to shrink with increasing $\gamma$.
The simplified 2D analysis by \citet{LatterOgilvie2006} doesn't present this $\gamma$ dependence in the ovestability condition caused by the shear viscosity, which means their overstability region always extends to high-$Q$ if the wavelength considered is sufficiently long. This discrepancy is believed to be due to their lack of a viscous heat modulation in the $A_\varv$ equation.

Since the term enclosed within square brackets in $\left( \frac{\partial \lambda}{\partial \gamma_s}\right)$ is usually positive, overstable conditions can be enhanced by minimising $\omega_0^2$ with respect to $k$; this is found to occur for

\begin{equation}
  k_{\mathrm{max}} = \frac{\pi G \Sigma_0}{c_s^2}.
\end{equation}
This value can then be used to calculate the critical value of $\gamma$ needed for overstability as a function of both $q$ and $Q$. The system is found to be overstable if

\begin{equation}
    \gamma < \frac{2-(2-q)^2 - \frac{4}{3}(2-q)\left(1-\frac{1}{Q^2}\right)}{q(2-q)},
\end{equation}
provided $Q\geq 1$.

In the non-SG limit the coefficient $(1-1/Q^2) \rightarrow 1$, reducing the overstability condition to

\begin{equation} \label{eq:gamma-crit-alphas}
  \gamma < \frac{16q -3q^2 - 14}{3q(2-q)},
\end{equation}
with the critical value being $\gamma_{\mathrm{crit}} \approx 1.444$ in the $q=3/2$ case.

The next step is to combine the contributions from different diffusivity types using Equation~\ref{eq:lambda-linear} to find the regions of the $k c_s/\Omega$--$Q$ plane where overstability would occur. We take the instance in which the bulk viscosity contribution is ignored; in this case, we find that the system would develop an overstability if 

\begin{equation}
  \frac{\gamma_t}{\gamma_s} < \frac{2\gamma \Omega^2 \left[(\gamma-1)q^2 + 2(2-\gamma)q -2 - \frac{2}{3}\omega_0^2/\Omega^2\right]}{k^2 c_s^2 (\gamma-1)},
\end{equation}
which, assuming $q=3/2$ and $\gamma=5/3$, simplifies to
\begin{equation}
  \frac{\gamma_t}{\gamma_s} < - \frac{5 \left[ 4k \left(kc_s^2 - 2c_s \kappa/Q\right) + \Omega^2\right]}{6 k^2 c_s^2} ,
\end{equation}
where the $\omega_0^2/\Omega^2$ factor has been expanded to obtain a relationship as a function of $k$. This highlights the stabilising effect played by thermal diffusion and cooling, with an overstability developing only if the ratio $\gamma_t/\gamma_s$ is below a critical value, which is dependent on the values of $k$ and $Q$ (as well as $q$ and $\gamma$). The cooling in particular plays a dominant role in the long-wavelength limit as its contribution to $\gamma_t$ is independent of $k$, while both shear and thermal diffusivities produce damping coefficients that are proportional to $k^2$. This hampers the triggering of overstability which, as seen in Figure~\ref{fig:overstability_alphas}, prefers the small $k$ limit, particularly for the non-SG case. An analysis of the $k\rightarrow 0$ limit, also taking into account the coupling between cooling timescale and shear viscosity given by the thermal balance (Eq.~\ref{eq:therm-eq-condition}), yields the following expression for the real part of the growth rate:

\begin{multline} \label{eq:k0-limit}
  \mathrm{Re}(\lambda) = \gamma_s \frac{\Omega^2}{6\kappa^2} \left[-28+4(8-3\gamma)q + 6(\gamma-1)q^2 \right. \\ \left. - 3(\gamma-1)^2q^2 \frac{1}{f_\mathrm{visc}} \right].
\end{multline}
%with the equality representing the disc being heated entirely by viscous dissipation.
From the expression above it is possible to infer that a non-SG overstability is indeed possible as long as the adiabatic index obeys $\gamma \lesssim 1.305$ (assuming $q=3/2$ still), with the threshold value $\gamma \simeq 1.305$ obtained when the disc is fully viscously heated ($f_\mathrm{visc}=1$). This represents a stricter constraint than that obtained for shear viscosity only (Equation~\ref{eq:gamma-crit-alphas}), again underlining the stabilising effect of $\gamma_t$. 

This is illustrated in Figure~\ref{fig:overstability_at_as}, where the area obeying $\omega_0^2<0$ has been ignored as any instability in that region would be of a dynamical nature. A range of cooling times satisfying thermal balance is explored, with the largest value chosen so that the flow is almost entirely heated by viscous dissipation. The Prandtl number is set to $\mathrm{Pr}=5$ with $\alpha_s=0.05$, for both $\gamma=1.3$ and $\gamma=1.4$. It is possible to notice that as the cooling is made more efficient the overstable area shrinks, confirming its stabilising role, particularly in the long-wavelength regime; indeed for $\gamma=1.4$, the system is found to be stable for all values of $k c_s/\Omega$ and $Q$ (for which $\omega_0^2>0$) for the shortest cooling time explored ($\tau_c=5\Omega^{-1}$). The $\gamma=1.3$ case, on the other hand, presents overstability for all cooling times explored, as predicted by Equation~\ref{eq:k0-limit}; non-SG or weak-SG conditions, overstability is also observed for $\gamma=1.3$ in the long-wavelength limit for the two longest cooling times analysed: $\tau_c=20\Omega^{-1}$ (green, full lines) and $\tau_c=29\Omega^{-1}$ (red, dot-dashed). Non-SG overstability, which requires wavelengths longer than $\sim 18H$ for $\gamma=1.3$, is on the other hand suppressed for $\gamma=1.4$, with overstable regions being contained to $Q\lesssim 2$. This is in agreement with the analytical prediction described above which found that a weak-SG/non-SG overstability in the $k\rightarrow 0$ limit could only be achieved if the value of the adiabatic index was below the threshold value $\gamma \simeq 1.305$. 

A general form for the largest overstable value of $Q$ attainable over all $k c_s/\Omega$ in the absence of the bulk viscosity contribution can be derived analytically and is found to be
\begin{multline} \label{eq:Qmax-over}
  \frac{1}{Q^2_\mathrm{max}} = \frac{\left[4\gamma + 3(\gamma-1) \mathrm{Pr}^{-1} \right]}{32(2-q)\gamma} \left[28-4(8-3\gamma)q - 6(\gamma-1)q^2 \right. \\ \left. + 3(\gamma-1)^2q^2 \frac{1}{f_\mathrm{visc}}\right].
\end{multline}
Assuming the sum of the first three terms enclosed in square brackets is positive (as otherwise the system might be overstable for any $Q$ and there would therefore not be a critical $Q$ value), $Q_\mathrm{max}$ is found to be an increasing function of $f_\mathrm{visc}$ and $\mathrm{Pr}$.
A particular example of Equation~\ref{eq:Qmax-over} is illustrated in Figure~\ref{fig:os_Q-vs-gamma}; this shows the overstability growth rates, maximised over $k$, as a function of the adiabatic index $\gamma$ and Toomre parameter $Q$ for $q=3/2$, $\tau_c=15\Omega^{-1}$, $\mathrm{Pr}=5$ and $\alpha=0.05$. While $\gamma$ values up to $\gamma\approx 1.6$ are overstable at $Q\sim 1$ for the given cooling timescale, the maximum $\gamma$ value needed for overstability gradually decreases to $\gamma\lesssim 1.25$ as the Toomre parameter reaches $Q\sim 5$. This is in agreement with the predicted maximum value of $\gamma$ that allows weak-SG/non-SG overstability ($\gamma \approx 1.305$), which is indicated by means of a dotted vertical line in the plot. Moreover, a dashed vertical line at $\gamma \approx 1.6$ represents the largest value of $\gamma$ allowed by thermal balance.

The introduction of the bulk viscosity in the analysis further complicates the overstability analysis, with the full form of the overstability criterion being
\begin{equation}
  \gamma_b < - \frac{\left[6k^2c_s^2\right] \gamma_t + \left[5\left( 4\omega_0^2-3\right) \Omega^2 \right] \gamma_s}{15 \omega_0^2},
\end{equation}
where the assumptions of $\gamma=5/3$ and $q=3/2$ have been made.

\subsection{Slow modes}
\begin{figure*}
  \centering
  \begin{minipage}{.99\textwidth}    
    \subfloat[]{\includegraphics[width=.47\columnwidth]{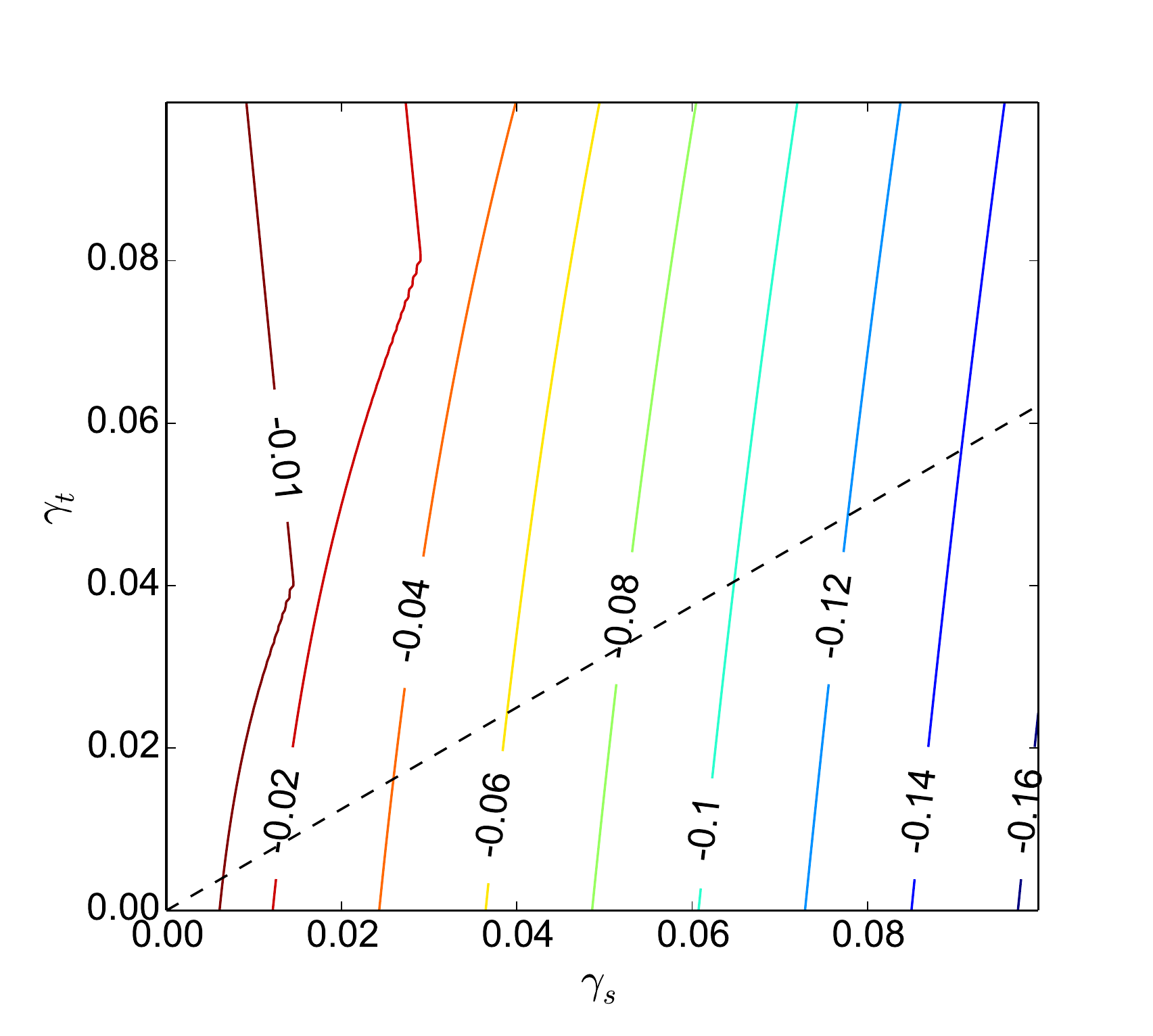} \label{fig:alphas_alphat_mode1}}
    \hspace*{.05\columnwidth}
    \subfloat[]{\includegraphics[width=.47\columnwidth]{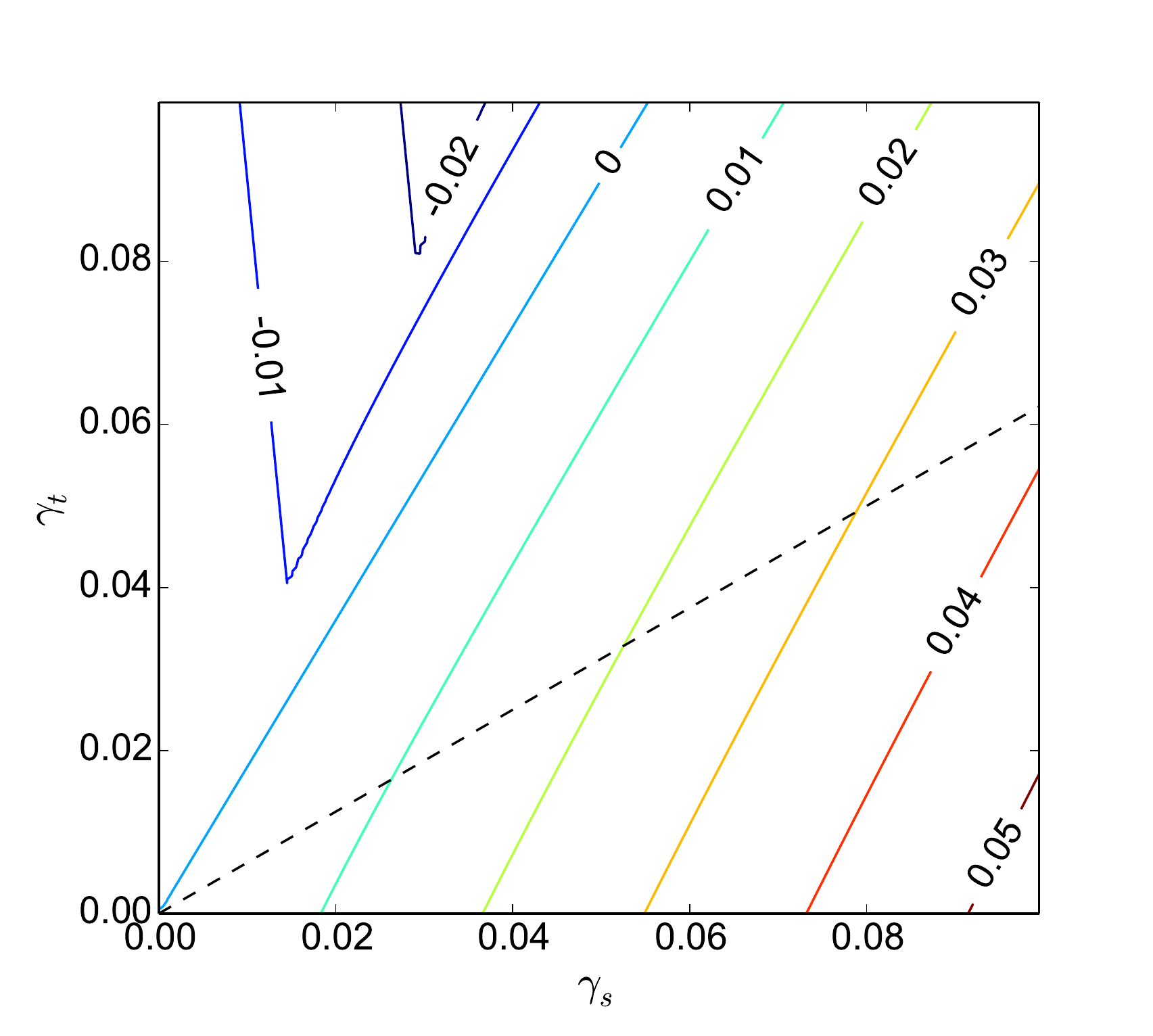} \label{fig:alphas_alphat_mode2}}
    \caption{Contour plot for the growth rates of the potential vorticity and entropy modes as functions of both $\gamma_s$ and $\gamma_t$, for $k c_s/\Omega=2$, $\gamma=5/3$ and $Q=1.2$. Both modes present non-linearities in their behaviour with an interference between the two modes observed for $\gamma_s \lesssim 0.03$, where the modes' growth rates are complex conjugates of one another. Value combinations of $\gamma_t$ and $\gamma_s$ below the dashed line do not obey thermal balance.}
    \label{fig:alphas_alphat_modes12}
  \end{minipage}
\end{figure*}

The analysis of the slow potential vorticity and entropy modes, having coinciding and degenerate solutions in the inviscid problem with no cooling or diffusion, requires a somewhat different approach from the regular perturbation method used for density wave modes, as their solutions are found to depend non-linearly with $\gamma_s$ and $\gamma_t$; this is exemplified in Figure~\ref{fig:alphas_alphat_modes12} for $\gamma=5/3$, $kc_s/\Omega=2$ and $Q=1.2$. The real parts of both modes' growth rates present non-linearities in their behaviour; interferences between the modes -- where their growth rates form a complex conjugate pair -- can also be observed for $\gamma_s \lesssim 0.03$. One of the two modes is also seen to be unstable in a sizeable part of the plot. Combinations of $\gamma_t$ and $\gamma_s$ values falling below the dashed line do not satisfy thermal balance (Equation~\ref{eq:therm-eq-condition}).

In order to gain a better understanding on the stability of these two modes, equations for the evolution of the structure in the specific entropy and potential vorticity (i.e. $\partial_t A_s$ and $\partial_t A_\zeta$) of the form

\begin{equation}
  \label{eq:Asdot}
  \partial_t A_s = c_1 A_s + c_2 A_\zeta,
\end{equation}
\begin{equation}
  \label{eq:Azetadot}
  \partial_t A_\zeta = c_3 A_s + c_4 A_\zeta,
\end{equation}
were analytically derived from Equations~\ref{eq:str-u}--~\ref{eq:str-e}, where $A_s$ and $A_\zeta$ are the dimensionless amplitudes of the axisymmetric structure in the respective quantities given by

\begin{equation}
  A_s = \frac{1}{\gamma} \left(A_e + A_h\right) ,
\end{equation}
\begin{equation}
  A_\zeta = \frac{k A_\varv}{(2-q)\Omega} - A_h,
\end{equation}
and $c_1$, $c_2$, $c_3$ and $c_4$ are coefficients which are independent of $A_\varv$, $A_h$ and $A_e$. The coefficients are found to be:

\begin{equation}
  c_1 = \frac{\gamma_t \left(c_s^2 k^2(\gamma-1)-\gamma\omega_0^2\right)+\gamma_s q \kappa^2 \gamma (\gamma-1)}{\gamma \omega_0^2},
\end{equation}
\begin{equation}
  c_2 = \frac{\kappa^2 (\gamma-1) \left[ \frac{\gamma_t c_s^2}{\gamma} + \frac{\gamma_s q}{k^2}\left(\kappa^2-\omega_0^2\right)\right]}{c_s^2 \omega_0^2},
\end{equation}
\begin{equation}
  c_3 = \frac{-4 \gamma_s c_s^2 k^2(q-1)\Omega^2}{\kappa^2 \omega_0^2}, 
\end{equation}
\begin{equation}
  c_4 = \frac{-\gamma_s \left(\omega_0^2 + 4 (q-1)\Omega^2\right)}{\omega_0^2}.
\end{equation}

Assuming the solutions have an exponential form, a generic quadratic equation for the growth rate $\lambda$ for the system described in Equations~\ref{eq:Asdot}--\ref{eq:Azetadot} can be simply derived:

\begin{equation}
  \label{eq:lambda-quadratic}
  \lambda^2 - (c_1 + c_4)\lambda + c_1 c_4 - c_2 c_3 = 0,
\end{equation}
with a generic solution being given by
\begin{equation}
  \label{eq:lambda_root_generic}
  \lambda = \frac{(c_1 + c_4)}{2} \pm \sqrt{\frac{(c_1-c_4)^2}{4} + c_2c_3}.
\end{equation}

The regions of the $k c_s/\Omega-Q$ space where the system is unstable to slow modes can be found by either looking for areas where $\mathrm{Re}(\lambda)>0$ or by applying a relevant stability condition. This was found in the Routh-Hurwitz stability criteria, which represent necessary and sufficient stability conditions for a linear time-invariant system with a polynomial characteristic equation. The required stability condition in the case of a generic second order polynomial of the form $x^2 + a_1 x + a_0=0$ is for all coefficients to satisfy $a_i>0$; in the particular instance of Equation~\ref{eq:lambda-quadratic}, this can be written as 
\begin{subequations}
\begin{align} 
  a_1 =&\,\, -(c_1+c_4)>0 \label{eq:inst-condition1},\\
  a_0 =&\,\, c_1 c_4 - c_2 c_3 > 0. \label{eq:inst-condition2}
\end{align}
\end{subequations}
Stability is achieved only if both of these conditions are satisfied.

\begin{figure*}
  \centering
  \begin{minipage}{.99\textwidth}    
    \subfloat[$\mathrm{Pr}=3$, $\alpha_s=0.05$]{\includegraphics[width=.47\columnwidth]{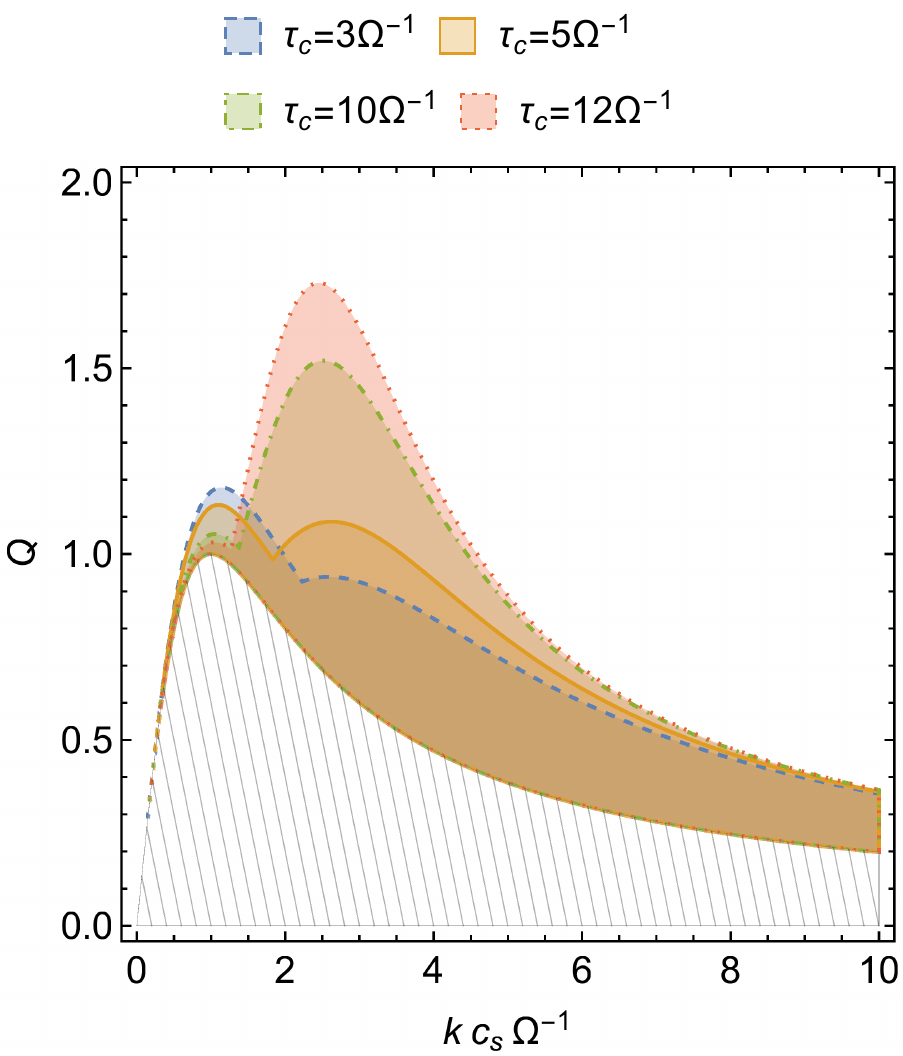}\label{fig:inst_cool_atas0-3_as5e-2_g53}}    
    \hspace*{.05\columnwidth}
    \subfloat[$\mathrm{Pr}=1$, $\alpha_s=0.05$]{\includegraphics[width=.47\columnwidth]{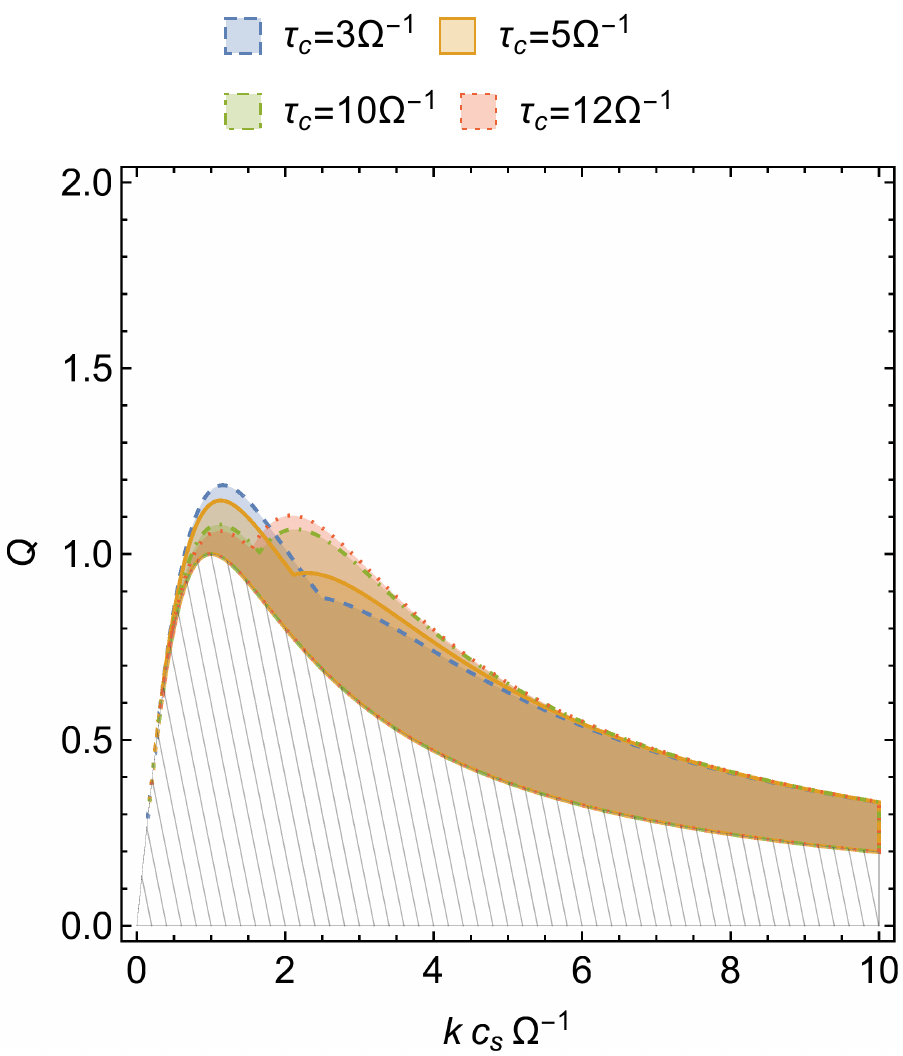}\label{fig:inst_cool_atas1_as5e-2_g53}}
    \caption{Instability regions in the parameter space given $\gamma=5/3$ and looking at the cooling times $\tau_c=3\Omega^{-1}$ (blue, dashed boundary), $\tau_c=5\Omega^{-1}$ (orange, full), $\tau_c=10\Omega^{-1}$ (green, dot-dashed) and $\tau_c=12\Omega^{-1}$ (red, dotted). The shear viscosity used is $\alpha_s=0.05$ and the Prandtl numbers \textbf{(a)} $\mathrm{Pr}=3$ and \textbf{(b)} $\mathrm{Pr}=1$. The instability features prominent peaks at $k c_s/\Omega \sim 2.5-3$ for $\mathrm{Pr}=3$, which are more noticeable for longer cooling times; these are quenched as the Prandtl number is decreased. Decreasing $\mathrm{Pr}$ also reduces the non-monotonic behaviour in the instability regions. The hatched area shows the region of the plane where $\omega_0^2<0$ and the system is therefore dynamically unstable to axisymmetric disturbances.}
    \label{fig:instability_cool_g53}
  \end{minipage}
\end{figure*}

\begin{figure*}
  \centering
  \begin{minipage}{.99\textwidth}    
    \subfloat[$\mathrm{Pr}=3$, $\alpha_s=0.05$]{\includegraphics[width=.47\columnwidth]{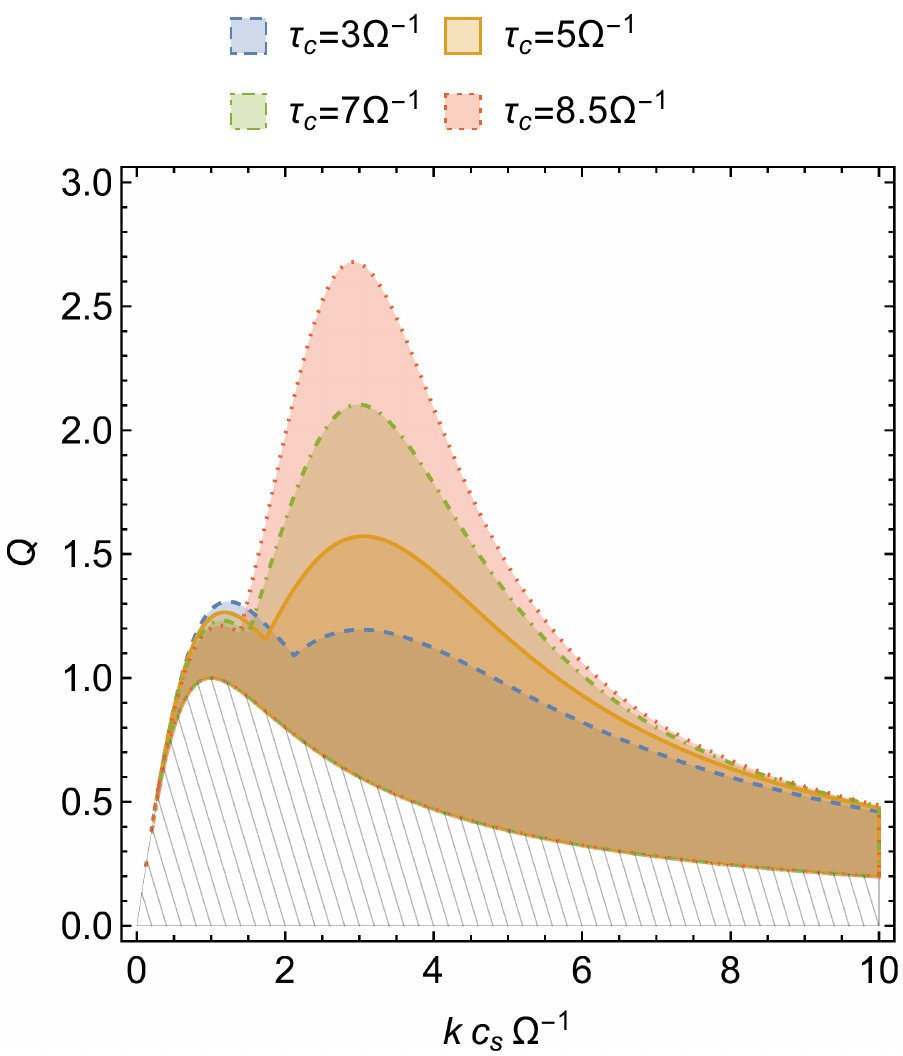}\label{fig:inst_cool_atas0-3_as5e-2_g2}}    
    \subfloat[$\mathrm{Pr}=1$, $\alpha_s=0.05$]{\includegraphics[width=.47\columnwidth]{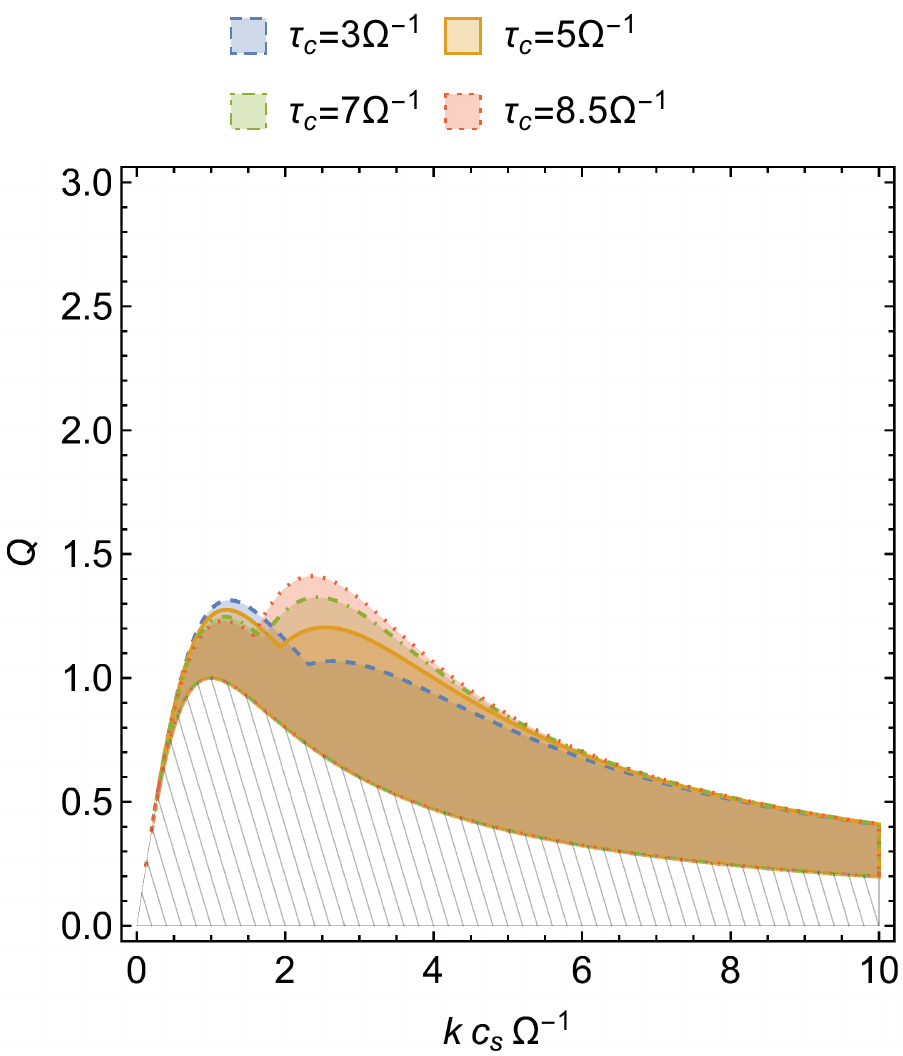}\label{fig:inst_cool_atas1_as5e-2_g2}}
    \caption{Similar analysis to Figure~\ref{fig:instability_cool_g53} but with $\gamma=2$ for a range of cooling times ($\tau_c=3\Omega^{-1}$ (blue, dashed boundary), $\tau_c=5\Omega^{-1}$ (orange, full), $\tau_c=7\Omega^{-1}$ (green, dot-dashed) and $\tau_c=8.5\Omega^{-1}$ (red, dotted)). The value of the shear viscosity is again $\alpha_s = 0.05$ throughout with the Prandtl number being \textbf{(a)} $\mathrm{Pr}=3$ and \textbf{(b)} $\mathrm{Pr}=1$. The larger value of $\gamma$ causes the instability regions to be larger than in the $\gamma=5/3$ case, particularly enhancing the peak at $k c_s/\Omega \sim 2.5-3$, which for $\mathrm{Pr}=3$ and $\tau_c=8.5\Omega^{-1}$ extends as far as $Q \approx 2.7$. This peak is however again quenched by decreasing the Prandtl number or by shortening the cooling timescale. The hatched area again shows the region of the plane where $\omega_0^2<0$.} 
    \label{fig:instability_cool_g2}
  \end{minipage}
\end{figure*}

\begin{figure*}
  \centering
  \begin{minipage}{.99\textwidth}    
    \subfloat[$\mathrm{Pr}=3$]{\includegraphics[width=.47\columnwidth]{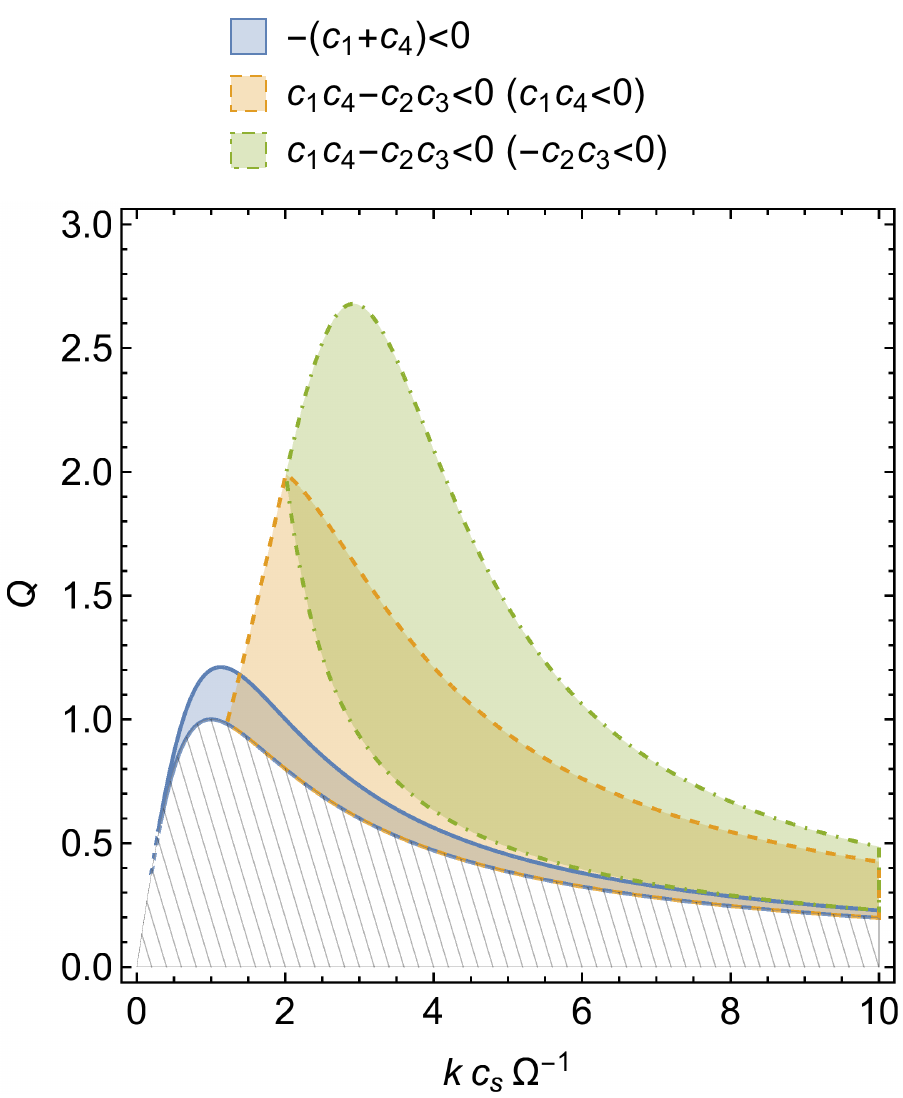}\label{fig:conditions_slow_atas0-3_as0-05_tc8-5_g2}}    
    \hspace*{.05\columnwidth}
    \subfloat[$\mathrm{Pr}=1$]{\includegraphics[width=.47\columnwidth]{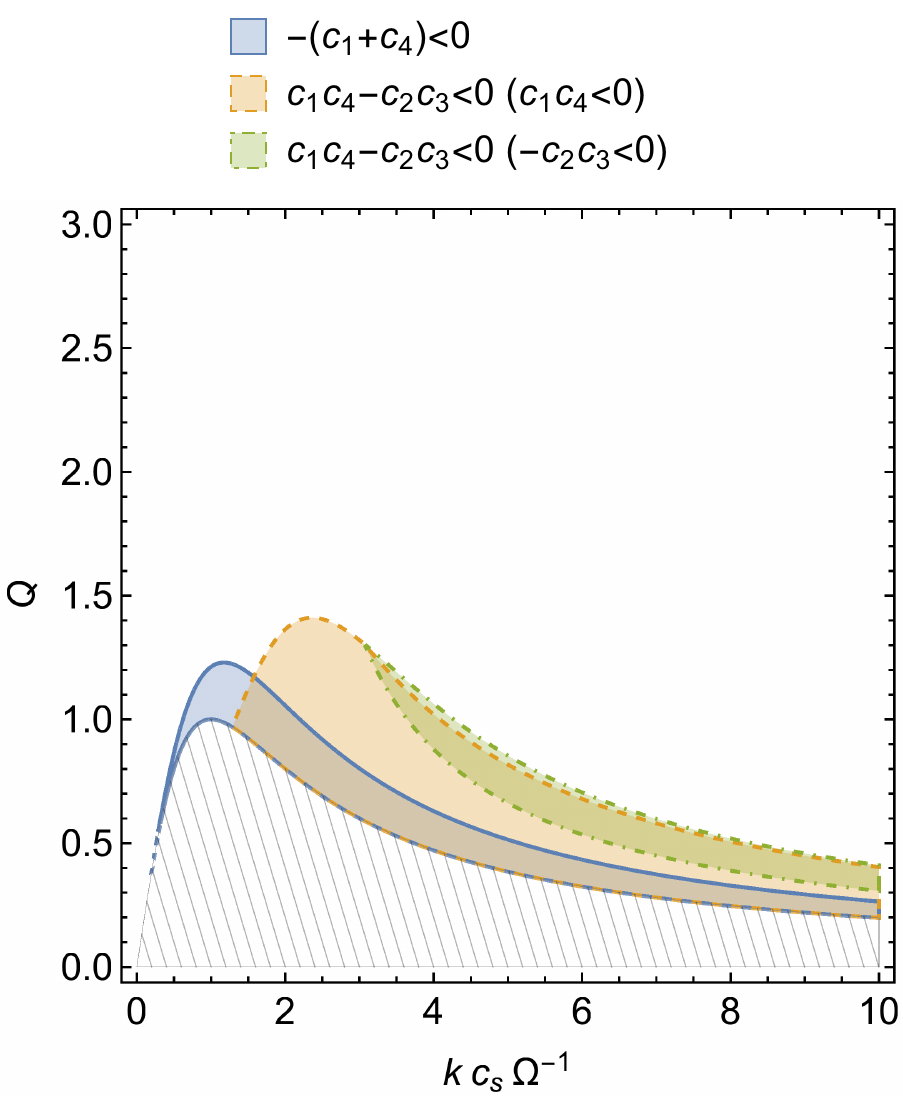}\label{fig:conditions_slow_atas1_as0-05_tc8-5_g2}}
    \caption{Analysis showing which of the two stability criteria is violated in the instability regions obtained in Figure~\ref{fig:instability_cool_g2} for $\gamma=2$, $\tau_c=8.5\Omega^{-1}$ and \textbf{(a):} $\mathrm{Pr}=3$ and \textbf{(b):} $\mathrm{Pr}=1$. Since $c_4<0$ as long as $\omega_0^2>0$ the first peak (blue, full line) is caused by an instability in the entropy ($c_1$), meaning it has a thermal nature. The second peak is due to the second instability criterion being fulfilled, with it being split among its two components. In \textbf{(a)} this is predominantly driven by the coupling term between entropy and PV (i.e. $-c_2c_3<0$, green dot-dashed); in \textbf{(b)} the decreased Prandtl number $\mathrm{Pr}$ quenches the coupling component almost completely with the $c_1c_4<0$ (yellow, dashed) mostly causing the instability, meaning this is driven by either PV or entropy. }
    \label{fig:instability_conditions}
  \end{minipage}
\end{figure*}

\begin{figure*}
  \centering
  \begin{minipage}{.99\textwidth}    
    \subfloat[$\mathrm{Pr}=3$]{\includegraphics[width=.47\columnwidth]{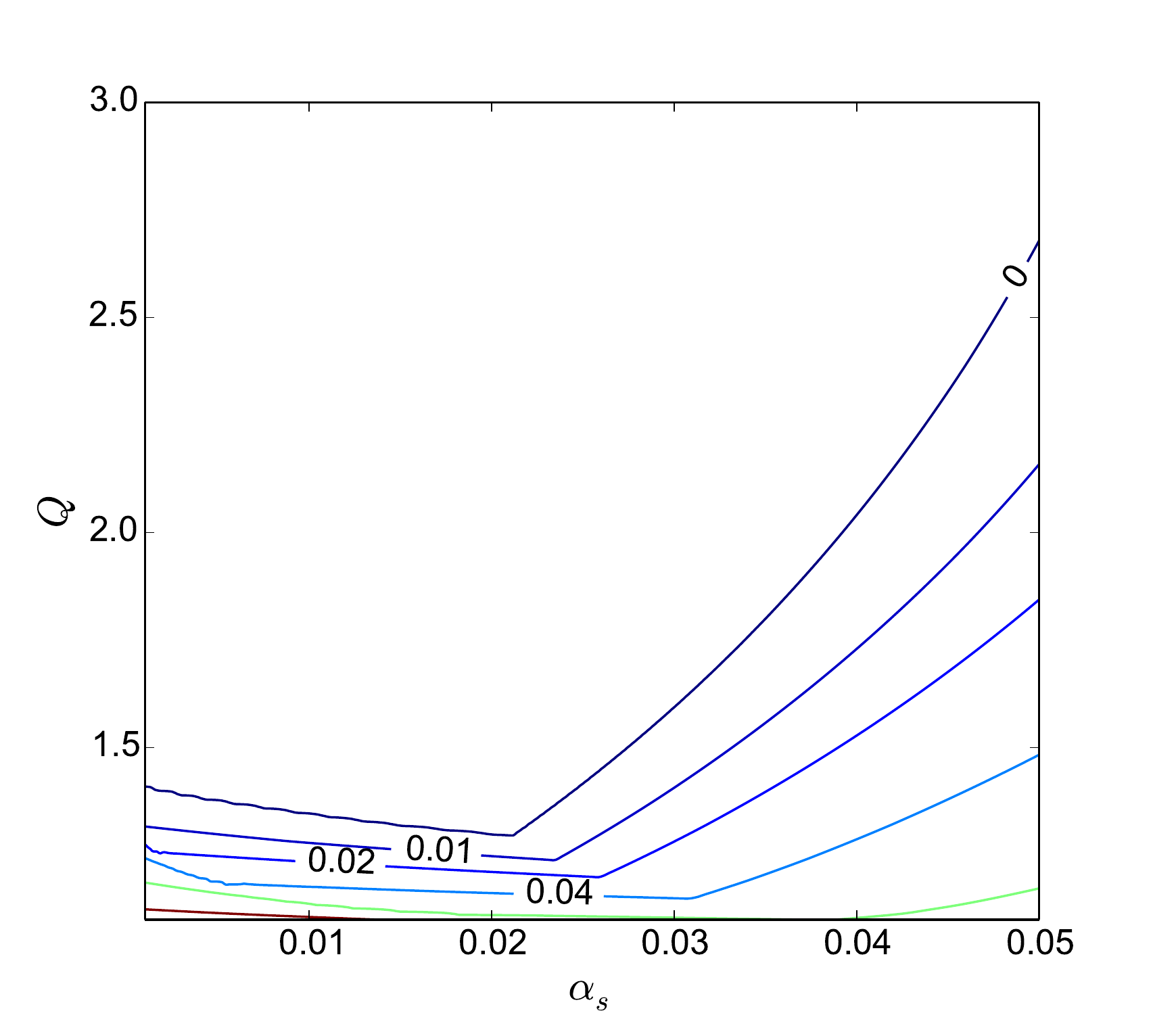}\label{fig:inst_alphas-vs-Q_atas1-3}}    
    \hspace*{.05\columnwidth}
    \subfloat[$\mathrm{Pr}=1$]{\includegraphics[width=.47\columnwidth]{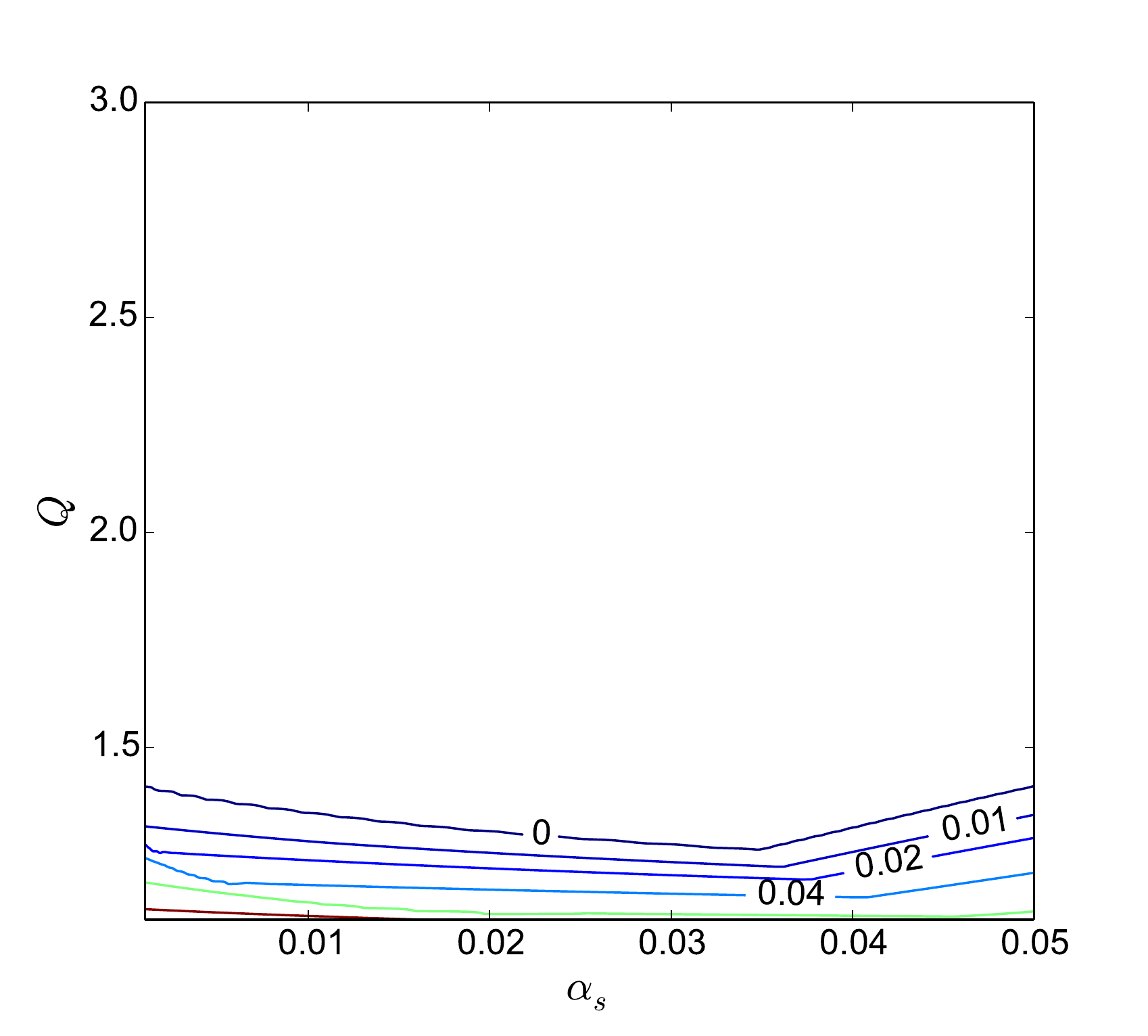}\label{fig:inst_alphas-vs-Q_atas1}}\\
    \caption{Instability growth rates maximised over $k$ as a function of the shear diffusivity $\alpha_s$ and Toomre parameter $Q$ for $q=3/2$, $\gamma=2$, $\tau_c=8.5\Omega^{-1}$ and a fixed Prandtl number of \textbf{(a):} $\mathrm{Pr}=3$ and \textbf{(b):} $\mathrm{Pr}=1$. Although the Prandtl number larger than unity remains a critical factor in boosting the instability, it is clear that the value of $\alpha_s$ is also of importance. For the smaller values of $\alpha_s$ plotted here, the disc is heated predominantly by external irradiation, while for the larger values of $\alpha_s$ it is mostly heated by viscous dissipation. For $\mathrm{Pr}=3$ no instability is seen above $Q\sim 1.4$ for $\alpha_s \lesssim 0.02$, although for larger values of $\alpha_s$ the instability spreads up to $Q \sim 2.7$; this points to the instability being boosted by a disc being viscously heated. The largest unstable $Q$ value is instead roughly constant in the $\mathrm{Pr}=1$ case. }
    \label{fig:inst_alphas-vs-Q}
  \end{minipage}
\end{figure*}

\begin{figure}
  \includegraphics[width=\columnwidth]{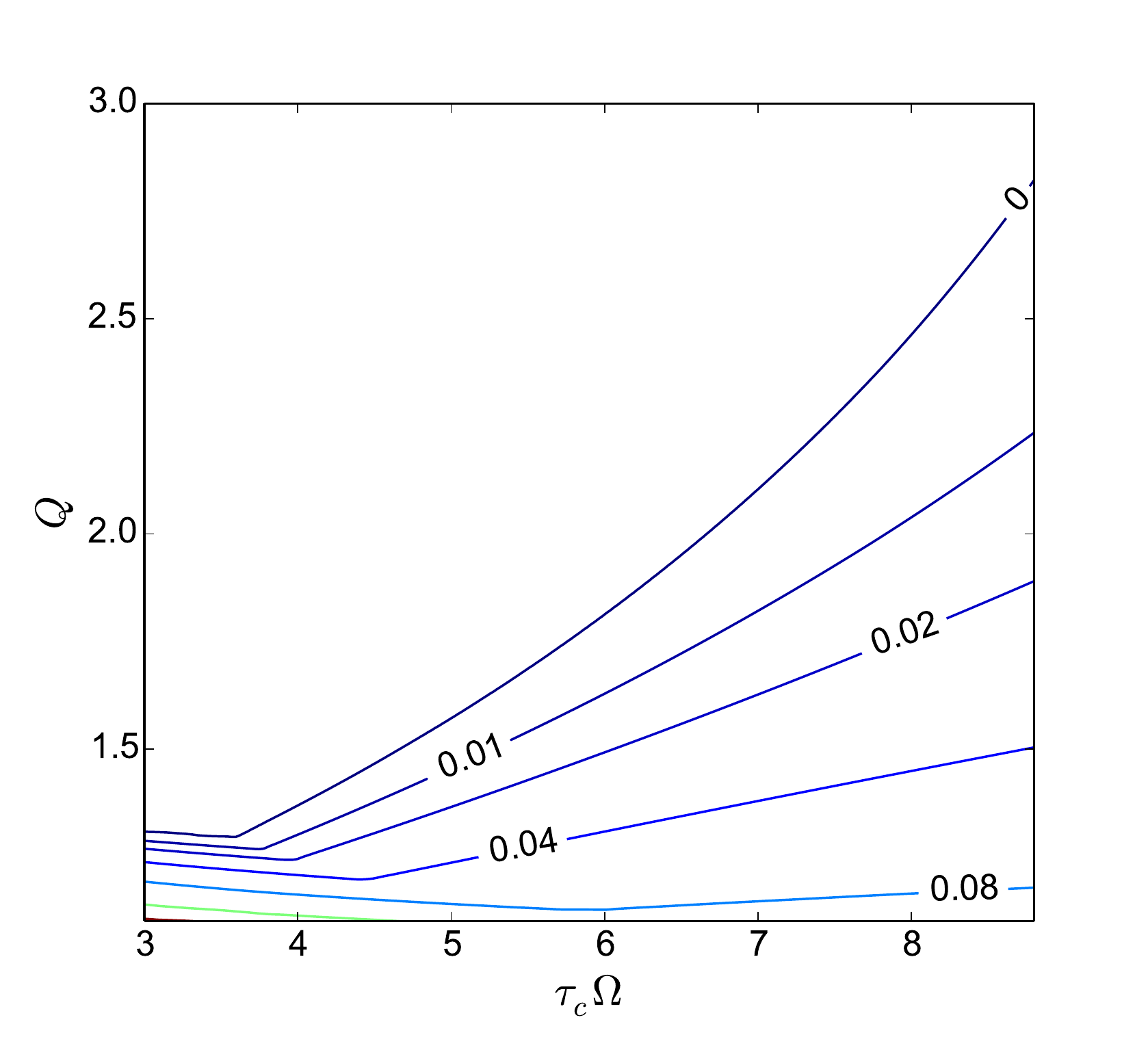}
  \caption{Instability growth rates, maximised over $k$, as a function of $Q$ and $\tau_c$ for $q=3/2$, $\gamma=2$, a Prandtl number of $\mathrm{Pr}=3$ and $\alpha_s=0.05$. For most of the $Q$ range, making the cooling timescale shorter has a stabilising effect on the system, as the peak at $k c_s/\Omega \sim 2.5-3$ in Figures~\ref{fig:instability_cool_g53} and~\ref{fig:instability_cool_g2} is quenched; for $Q\lesssim 1.25$ the trend however reverses for very short cooling times which is due to the peak at $k c_s/\Omega \sim 1$, having a thermal nature, being boosted.}
  \label{fig:instability_slow_Q-vs-t}
\end{figure}

If the coupling coefficients $c_2$ and $c_3$ are negligible compared to $c_1$ and $c_4$, entropy and PV evolve independently from each other, with $c_1$ and $c_4$ representing the two quantities' respective growth rates. Such is the case in both long-wavelength (i.e. $kc_s/\Omega \rightarrow 0$) and short-wavelength (i.e. $kc_s/\Omega \rightarrow \infty$) limits, the former being stable according to the classical approach. In these cases the product of $c_1$ and $c_4$ -- both coefficients being negative -- dominates over the coupling product term $c_2c_3$; this means that both Routh-Hurwitz stability criteria are satisfied and the system is stable. The analysis presented in this paper focuses on the stability of the intermediate $kc_s/\Omega$ range, instead; this is somewhat more difficult to predict analytically as $c_2$ and $c_3$ are no longer negligible, meaning PV and entropy are coupled. This also implies that, should both $c_1$ and $c_4$ be negative under certain conditions, the system can nevertheless still be unstable by violating the $c_1c_4-c_2c_3>0$ condition.

It's worth pointing out that the properties of the model used do affect the stability of the flow; should the $+q\Omega \gamma_s k^{-1} A_h$ term in Equation~\ref{eq:str-v} -- which arises from the dynamic viscosities being linear functions of $\Sigma$ -- be removed, the system would then be unstable to secular gravitational instability \citep{Willerding1992, Gammie1996}. This occurs in the limit $k c_s/\Omega \rightarrow 0$ in systems which are marginally stable according to Equation~\ref{eq:visc-inst-criterion}. However in the case analysed in this work, the system is stable to the onset of secular gravitational instability. 

Figure~\ref{fig:instability_cool_g53} illustrates the regions in the $k c_s^2/\Omega$--$Q$ plane where either (or both) of the stability conditions is not satisfied and the system is therefore unstable; the same instability regions are also obtained when looking for parts of the plane where $\mathrm{Re}(\lambda)>0$, therefore validating the instability criteria used. The analysis is carried out for a range of cooling times satisfying thermal balance (where again the largest value is such that $e_{\mathrm{irr}} \simeq 0$) with $q=3/2$ and $\gamma=5/3$ and for $\mathrm{Pr}=3$ (Figure~\ref{fig:inst_cool_atas0-3_as5e-2_g53}) and $\mathrm{Pr}=1$ (Figure~\ref{fig:inst_cool_atas1_as5e-2_g53}). The value of the shear viscosity is kept at $\alpha_s=0.05$ throughout.
Non-monotonic behaviour in the instability regions is observed thanks to a peak at $k c_s/\Omega \sim 2.5-3$, which is most
 prominent for $\mathrm{Pr}=3$ but is quenched as the Prandtl number decreases to unity. The overall region of instability also shrinks with decreasing $\mathrm{Pr}$, highlighting the stabilising effect of the thermal diffusion. A short cooling time seems to lightly boost instability at $k c_s/\Omega \sim 1$, but at the same time it appears to dampen the instability at $k c_s/\Omega \sim 2.5-3$.
All instability regions seem to prefer intermediate $kc_s/\Omega$ values, ensuring the instability is again very relevant to the stability of zonal flows. The effect the value of the cooling timescale has on the stability of the system appears to wane with decreasing Prandtl number, with the $\mathrm{Pr}=1$ case presenting a reduced difference between the $\tau_c=3\Omega^{-1}$ and $\tau_c=12\Omega^{-1}$ cases.

Figure~\ref{fig:instability_cool_g2} represents a similar analysis to Figure~\ref{fig:instability_cool_g53}, but this time with the adiabatic index set to\footnote{Although the value $\gamma=2$ bears questionable physical relevance, this has regularly been adopted in works of self-gravitating accretion discs since the seminal analysis by \citet{Gammie2001}. It is therefore useful in comparing our results to the relevant literature.} $\gamma=2$. The increased value of $\gamma$ causes a boost in both peaks compared to the $\gamma=5/3$ case, particularly the one located at $k c_s/\Omega \sim 2.5-3$. Once again, this latter peak is suppressed as the Prandtl number is decreased with the non-monotonic behaviour mostly suppressed for $\mathrm{Pr}=1$. Also, as seen in Figure~\ref{fig:instability_cool_g53}, the use of an effective cooling has the effect of boosting the first peak (the one at $k c_s/\Omega\sim1$), while quenching the second one. 

The nature of the instability region is explored in Figure~\ref{fig:instability_conditions} with $\gamma=2$, $\tau_c=8.5\Omega^{-1}$ and $\mathrm{Pr}=3$ and $\mathrm{Pr}=1$ in Figures~\ref{fig:conditions_slow_atas0-3_as0-05_tc8-5_g2} and~\ref{fig:conditions_slow_atas1_as0-05_tc8-5_g2}, respectively. The total unstable area is divided into the regions where each of the stability conditions given in Equations~\ref{eq:inst-condition1}--~\ref{eq:inst-condition2} is violated. The first peak, located at $k c_s/\Omega \sim 1$, is due to the $-(c_1+c_4)>0$ stability condition being violated and it therefore represents, as suggested by Equation~\ref{eq:lambda_root_generic}, an oscillatory instability. As $c_4<0$ (assuming $\omega_0^2>0$), regardless of the values of $Q$ or $k c_s/\Omega$, the unstable contribution must come from $c_1$, meaning that part of the region is caused by an instability in the entropy; this is therefore a thermal instability. On the other hand the second peak, found at $k c_s/\Omega \sim 2.5-3$, is triggered by the second condition not being fulfilled (i.e. we therefore have $c_1c_4-c_2c_3 <0$), implying the instability here has a non-oscillatory behaviour; this peak is therefore due either to the action of entropy or potential vorticity (orange, dashed region; $c_1c_4<0$), or to their coupling (green, dot-dashed region, $-c_2c_3<0$), as seen in Equations~\ref{eq:Asdot}--\ref{eq:Azetadot}. The comparison between Figures~\ref{fig:conditions_slow_atas0-3_as0-05_tc8-5_g2} and~\ref{fig:conditions_slow_atas1_as0-05_tc8-5_g2} shows that decreasing the Prandtl number results in the quenching of the coupling's destabilising effect, with said coupling mostly driving the instability at $kc_s/\Omega \sim 2.5-3$ for $\mathrm{Pr}=3$ but it being largely suppressed in the $\mathrm{Pr}=1$ case. A small boost of the entropy-driven instability is also observed upon decreasing $\mathrm{Pr}$.

Figure~\ref{fig:inst_alphas-vs-Q} shows the growth rates of the instability region, which have been maximised over $k$, as a function of $\alpha_s$ and $Q$; these are obtained for $\gamma=2$, $\tau_c=8.5\Omega^{-1}$ and a fixed Prandtl number of $\mathrm{Pr}=3$ (Figure~\ref{fig:inst_alphas-vs-Q_atas1-3}) and $\mathrm{Pr}=1$ (Figure~\ref{fig:inst_alphas-vs-Q_atas1}). All values of $\alpha_s$ used are allowed by thermal balance for the given cooling timescale, with small $\alpha_s$ values indicating the disc is predominantly heated by external irradiation while the maximum explored value of $\alpha_s=0.05$ means the disc is almost completely heated by viscous effects. The plot shows that while the value of the Prandtl number is of importance for the stability of the system, the value of $\alpha_s$ -- and therefore the source of internal energy -- also affects the maximum value of $Q$ at which the instability is observed. Indeed for $\mathrm{Pr}=3$ the system is unstable up to $Q\sim 2.7$ for $\alpha_s=0.05$ (viscously heated disc), but only up to $Q \sim 1.4$ when $\alpha_s \lesssim 0.02$ (external irradation contributing at least as much as viscous effects). The $\mathrm{Pr}=1$ case, on the other hand, presents little variation in $Q$ over the diffusivity range, although a similar qualitative behaviour is observed.

The dependence of the $k$-maximised growth rates on $Q$ and the cooling time $\tau_c$ for $\alpha_s=0.05$ and $\mathrm{Pr}=3$ is instead explored in Figure~\ref{fig:instability_slow_Q-vs-t}. This shows that while for most of the $Q$ range shortening the cooling time has a stabilising effect on the system, due to the peak at $k c_s/\Omega \sim 2.5-3$ being quenched as seen in Figures~\ref{fig:instability_cool_g53} and~\ref{fig:instability_cool_g2}, the situation is reversed for $Q\lesssim 1.25$. This is caused by the peak observed at $k c_s/\Omega \sim 1$, which possesses a thermal nature as seen in Figure~\ref{fig:instability_conditions}, being instead boosted by efficient cooling. 

\section{Conclusions} \label{sec:conclusions}
We carried out an analytical calculation on the evolution of a viscous and compressible self-gravitating Keplerian disc having a constant cooling timescale and horizontal thermal diffusion with an axisymmetric structure present in the analysed quantities. The analysis took into account all solutions of the problem: both the density wave modes and the potential vorticity and entropy slow modes.  

While the solutions to the system are well known in the inviscid case with no cooling or thermal diffusion, the introduction of three types of diffusivity (bulk and shear viscosities and thermal diffusion) and cooling created a non-trivial problem in pinpointing whether they would have a stabilising or destabilising effect on the system. A simplification can be made for the density wave modes, as their growth rates are found to be a linear function of each type of diffusivity used (regular perturbation method); this allowed us to individually derive the contribution from each diffusivity type to the final growth rate. These contributions can then be summed together to establish the actual growth rate of the modes. While the bulk and thermal diffusivities were found to always have a stabilising effect, the situation was somewhat more complex for the shear viscosity. Ignoring the contribution made by the bulk viscosity, the system was found to be overstable for intermediate and long wavelengths for Toomre parameter values of $Q\lesssim2$, although a weak-SG/non-SG overstability was also detected in the long-wavelength regime for inefficient cooling as long as the adiabatic index $\gamma \lesssim 1.305$. In the case of $\gamma=1.3$ the system is overstable for non-SG conditions for wavelengths longer than roughly $18H$. These results appear consistent with those by \citet{LatterOgilvie2006} in the simplified 2D version of their calculation, although their work did not present any $\gamma$-dependence due to the lack of thermal heating modulations in the azimuthal velocity equation. The $k$-maximised growth rates for overstability regions were plotted as a function of adiabatic index and $Q$; while a sizeable range of $\gamma$ values presented overstability for $Q\sim 1$, this gradually reduced as $Q$ was increased. Only values obeying $\gamma \lesssim 1.25$ were found to be overstable in weak self-gravitating conditions for $Q\sim 5$, which is in agreement with the predicted threshold of $\gamma \lesssim 1.305$. Overstability criteria for shear and thermal diffusivities only and for all three diffusivity types were also derived, which highlight the stabilising effect of thermal diffusivity in the weak-SG regime. 

The situation was more complex for the entropy and potential vorticity slow modes as their degenerate solutions in the inviscid case with no cooling were found not to follow the regular pertubation method. In order to obtain their growth rates, equations for the evolution of the axisymmetric structure in these two quantities -- which only depended on the structure's amplitude in the entropy and PV themselves -- were derived. The Routh-Hurwitz stability criteria, representing the conditions for which a linear time-invariant system with a polynomial characteristic equation is stable, were applied to the generic solution to these equations. The long- and short-wavelength limits, which are stable according to the classical stability analysis, were likewise found to be stable. Nevertheless, the flow was found to be unstable in the intermediate wavelength regime, in a clear extension to the classical approach. This instability was found to be aided by considering higher values of the adiabatic index and of the Prandtl number and by decreasing the values of the Toomre parameter, although it was also of importance whether the disc was heated by external irradiation or viscous effects. Efficient cooling, on the other hand, was found to have an overall stabilising effect on the instability as long as $Q\gtrsim 1.25$. It is believed that this kind of instability -- due to its tendency to operate at intermediate wavelengths -- might result, in the appropriate conditions, in the formation of zonal flows; these might themselves be unstable, potentially giving rise to vortices in the flow. Further work is however required to obtain a more detailed link between the instability and the potential development of zonal flows. 

\section*{Acknowledgements}
We would like to thank the reviewer for providing a constructive set of comments.
The research was conducted thanks to the funding received by the Science \& Technology Facilities Council (STFC).

%%%%%%%%%%%%%%%%%%%%%%%%%%%%%%%%%%%%%%%%%%%%%%%%%%



\bibliographystyle{mnras}
\bibliography{diffusive_mathematica_paper.bib} % if your bibtex file is called example.bib

%%%%%%%%%%%%%%%%%%%%%%%%%%%%%%%%%%%%%%%%%%%%%%%%%%

%%%%%%%%%%%%%%%%% APPENDICES %%%%%%%%%%%%%%%%%%%%%

% Don't change these lines
\bsp	% typesetting comment
\label{lastpage}
\end{document}